\documentclass[aps,showpacs,preprintnumbers,amsmath,amssymb,nofootinbib]{revtex4}

\usepackage{graphicx}
\usepackage{color}

\def \be  {\begin{equation}}
\def \ee  {\end{equation}}
\def \ee  {\end{equation}}
\def \bea {\begin{eqnarray}}
\def \eea {\end{eqnarray}}

\def\be {\begin{equation}}
\def\ee {\end{equation}}
\def\bea {\begin{eqnarray}}
\def\eea {\end{eqnarray}}
\def\bc {\begin{center}}
\def\ec {\end{center}}
\def\bfg {\begin{figure}}
\def\efg {\end{figure}}
\def\bi {\begin{itemize}}
\def\ei {\end{itemize}}

\def\beq{\begin{equation}}
\def\eeq{\end{equation}}
\def\br{\begin{eqnarray}}
\def\er{\end{eqnarray}}
\newcommand{\eel}[1] {\label{#1}\end{equation}}
\begin{document}

\def\appendixa{
\vskip 1cm
 {\bf Appendix A: Wave function with the Clebsch-Gordon coefficient}
 \vskip 1cm
 \par
 \setcounter{equation}{0}
 \def\theequation{A.\arabic{equation}}
 }
%
 \def\appendixb{
\vskip 1cm
 {\bf Appendix B: The Derivation of $OBEP$ through the exchange of two mesons}
 \vskip 1cm
 \par
 \setcounter{equation}{0}
 \def\theequation{B.\arabic{equation}}
 }
%
\def\appendixc{
\vskip 1cm
 {\bf Appendix c: The treatment of pseudoscalar meson within our potential}
 \vskip 1cm
 \par
 \setcounter{equation}{0}
 \def\theequation{C.\arabic{equation}}
 }
%
\preprint{2019-09}
\title{The Ground State Calculations for Some Nuclei by Mesonic Potential of Nucleon-Nucleon Interaction}

\author{K.M.Hanna}
\affiliation{Mathematics :: Theoretical Physics Department, Atomic Energy Authority, Cairo, Egypt}

\author{L. I. Abou-Salem}
\email{loutfy.Abousalem@fsc.bu.edu.eg}
\affiliation{Physics Department, Faculty of Science, Benha University, Benha 13518, Egypt}

\author{SH.M.Sewailem}
\email{Sh_m_sw@yahoo.com}
\affiliation{Mathematics :: Theoretical Physics Department, Atomic Energy Authority, Cairo, Egypt}

\author{Asmaa.G.Shalaby}
\email{asmaa.shalaby@fsc.bu.edu.eg}
\affiliation{Physics Department, Faculty of Science, Benha University, Benha 13518, Egypt}

\author{R.Hussien}
\email{rabab.hussien216@gmail.com}
\affiliation{Mathematics :: Theoretical Physics Department, Atomic Energy Authority, Cairo, Egypt}

\begin{abstract}
\par The interaction of Nucleon-Nucleon $(NN)$ has certain physical characteristics, which indicated by nucleon, and meson degrees of freedom. The main purpose of this work is calculating the ground state energies of $^{2}_{1}H$ and $^{4}_{2}He$ through the two-body system with the exchange of mesons $(\pi,\sigma,\omega)$ that mediated between two nucleons. This paper investigates NN interaction based on the quasi-relativistic decoupled Dirac equation and self-consistent Hartree-Fock formulation. We construct one-boson exchange potential (OBEP) model, where each nucleon is treated as a Dirac particle and acts as a source of pseudoscalar, scalar, and vector fields. The potential in the present work is analytically derived with two static functions of meson, the single particle energy dependent(SPED) and generalized Yukawa (GY) functions, the parameters used in meson functions are just published ones (mass, coupling constant,and cut off parameters). The theoretical results are compared to other theoretical models and their corresponding experimental data, one can see that the SPED function gives more satisfied agreement than the GY function in case of the considered nuclei.
\end{abstract}
\maketitle
keywords:Nucleon-Nucleon interaction, Dirac equation, Hartree-Fock formulation, One Boson Exchange.

\section{Introduction}
\par
One of the aims of nuclear structure theory is to derive the ground state properties. Such properties are related to the constituents of matter, which are represented in the physics of elementary particles with their characteristics (electric charge, mass, spin,...) and how each particle interacts with others \cite{griffiths}. Yukawa (1934) introduced an assumption of some sort of field to be the reason of attraction between proton and neutron. This field is quantized, characterized by the force of the short range, and its mass equals $300$ times of electron called Yukawa particle (meson). Meson is a Greek word, means intermediate and this is the right description for meson which transmits the nuclear force between hadrons, meson can participate in weak,strong and electromagnetic interactions with a net electric charge. All mesons are unstable and their lifetimes reach to hundredths of microseconds. Each meson characterized by quantum numbers, principle number $(n=0,1,..)$, orbital angular momentum ($l=n-1$; indicates to the orbiting of quarks around each other), magnetic number $(m=-l,...,l)$, and spin ($s= 0$ for singlet state, $1$ for triplet state). The description of quantum numbers can be illustrated by using the nuclear shell model \cite{vallieres}.
\par
The interaction between each nucleon with all other nucleons generates an average potential field where each nucleon moves. The rules of Pauli exclusion principle govern the occupation of orbital quantum states in the shell-model and postulate that under the meson-exchange between two nucleons, the wave function is totally antisymmetrical product wave function. Nucleons interaction forms a potential characterized by its dependence on position named with nuclear mean-field. There is an ability to calculate mean-field potential for electrons or in nuclei, the calculation methods are very similar, but the interactions are different. The $NN$-interaction is a fundamental problem in nuclear physics. It had a variant success in describing the nuclear properties, this determination ranged from the empirical picture to fit the experimental data, to derive it microscopically from the bare $NN$ potential. Thus, there is no unique $NN$ potential to be the start point \cite{kamal85,scot2002,sahu14}.\\
A microscopic description was provided in nuclear models to include the elementary interaction between nucleons. The original attempts to find the fundamental theory of nuclear forces \cite{yukawa35,james17,taketani52,brueckner53,signell57} were not so successful. The reason for their failure was the pion dynamics which has been restricted by the chiral symmetry \cite{brown91,naghdi14}. The quantum description and field methods were included in making the potential structure such as Partovi-lomon model \cite{partovi70}, Stony Brook group \cite{jackson75}, Paris-group \cite{lacombe73}, Nijmegen-group \cite{nagels78} and Bonn-group potentials \cite{naghdi14}. The successful theoretical models were based on (OBEP), including one-meson exchange and multi-mesons exchange, plus short range phenomenology.  Inside a nucleus, there is a fact that nucleons move quasi-independently from one to another which achieves the concept of Nuclear Mean Field (NMF), this fact relies on Hartree-Fock interaction.\\
The microscopic description of degrees of freedom related to nucleon and meson have to depend on a relativistic quantum field to include the full structure of the medium (spin structure) which is associated with the fermions field of Dirac equation \cite{bouyssy87}, and the bound state energies. This can be found by solving the Dirac equation which leads to investigating the ground state energies of nuclei which is ensured by the calculation of the NMF potential with Dirac-Hartree-Fock \cite{jaminon81}.
\par
It is known that the $NN$ interaction can be distinguished into three parts, the first part is the Long-Range at $r\geq 2$ fm originated from pseudoscalar mesons, the second part is the Medium-Range at $1fm\leq r \leq 2fm$, which mainly comes from the exchange of scalar-meson($\sigma$ which is a fictitious scalar meson responsible for attraction), the third part is the Short-Range at $ r\leq 1fm$, from the vector-meson $(\rho,\omega,...)$ exchange. In order to have a potential of $NN$ interaction, there were many models that serve this point \cite{reid68,day81,stoks94,lagaris81,wiringa95,lassila62,hamada62}.
After little development in nuclear properties, the dominant part of interaction is central, having a strong repulsion at short range $(r\leq 0.7fm)$, and attraction force at intermediate range $(> 1 fm)$. There is a cancelation of major static effects between vector and scalar mesons to maintain the stability of nucleus\cite{watari67}.
\par
Now, a variant number of pseudoscalar, scalar, and vector mesons are found, the vast advance of $OBEP$ models  related to $NN-$interaction not only for free parameter reduction, but also in the accuracy and fitting them with experimental data \cite{Machleidt87,naghdi14}. The development of quantum field theory and boson field Lagrangian by Heisenberg, Pauli, Dirac, and Rosenfeld in (1930), allow the meson field coordinates to depend on themselves by Yukawa in (1935). Firstly, Yukawa suggested the conjunction of scalar field coordinate of mesons, and then extended to include vector fields by Proca (1936), Kemmer (1938) embraced the pseudoscalar, axial vector and antisymmetric tensor. Till now, there are a large number of modifications in the vector-scalar combinations as well as pseudoscalar and pseudovector mesons.
\par
Present work represents a motivated model of determining the ground state for deuteron$(^{2}H)$ and helium $(^{4}He)$  based on $NN-$interaction, the potential related to $OBEP$ with the exchange of pseudoscalar meson $(\pi)$, scalar meson $(\sigma)$ and vector meson $(\omega)$. This potential is derived analytically with two static functions of mesons, it relies on Dirac-Hartree-Fock equation. Then, we compare the obtained theoretical results with others and their corresponding experimental data.
\par
This paper arranged as follows. Section \textbf{II}, is devoted to explain the theoretical analysis in details, with three subsections, A, B and C. Subsection \textbf{A}, refers to how this model use Hartree-Fock equation with the Dirac Hamiltonian and how it deals with the wave functions. Subsection \textbf{B} is related to the mathematical treatment of each term in the Hamiltonian equation. Subsection \textbf{C}, represents the potential of our model. Section \textbf{III}, represents the results of the potential and groundstate energy for the selected nuclei. Finally, section \textbf{IV}, is the conclusion.
\section{Theoretical Analysis}
\par
There are several models which determine the structure and properties of the nucleus through the strong force between nucleons in the nucleus \cite{zahra16, haidenbauer}. The states of nucleus are bounded due to this strong force. The nature of nucleons interactions can be described by studying it as two-body problem. The general wave equation used in such models has the form.
 \begin{equation}
\hat{ H}|\Psi\rangle = E |\Psi\rangle
\end{equation}
Where $\hat{H}$ represents the general Hamiltonian operator, and $E$ is the eigen energy. We studied the interaction through two-body via $OBEP$ between two fermions (nucleons) so, the convenient representation of the energy is the relativistic form of Dirac equation.
\newpage
Thus, the accurate interaction of the nuclear system can be described by Dirac Hamiltonian which include all fermions interaction and given by\cite{miller72,meibner2005,arias2001}
 \begin{equation}\label{1}
\hat{H}=\sum_{i}^{A}c\vec{\alpha_{i}}.\vec{p_{i}}+(\beta_{i}-I)m_{i}c^2-\hat{T}+\frac{1}{2}\sum_{i\neq j}^{A}V_{ij}
\end{equation}
Since $I$ is the unit matrix, $\vec{\alpha}$ and $\beta$ are $(4\times 4)$ Dirac matrices, $m_{i}$ is the nucleon mass, $\vec{p_{i}}$ is the momentum of the system, $\hat{T}$ is kinetic energy operator and $V_{ij}$ is the potential energy between fermions' pairs and we ignore three and many body interactions in present work. The total kinetic energy of the nucleon equals the total energy subtracted from the rest mass energy\cite{jaminon80}.
\begin{equation}\label{r}
\hat{T} = E - Mc^{2}
\end{equation}
Where $M = A m_{i}$, with $A$ is the number of nucleons and $E$ is the total relativistic energy which has the form,
\begin{equation}
 E = M c^2\sqrt{1+\frac{p^2}{M^{2}c^{2}}}
\end{equation}
The kinetic energy can be decomposed into two contributions, the first one is the relative space contribution $\hat{T_{r}}$, and the other is the center of mass contribution $\hat{T}_{cm}$ \cite{neff15,takayuki2009}.
\begin{eqnarray}\label{kinetic}
\nonumber \hat{T}&=& \sum_{i}\hat{T_{r}}-\hat{T}_{cm}\\
     &=& \frac{(\sum _{i}^{A}p_{i})^{2}}{2m A}-\frac{\sum _{i}^{A}p_{i}^2}{2m A}
\end{eqnarray}

The second part of Eq. (\ref{kinetic}) can be neglected. This neglecting the center of mass term $\left(\sum_{i}^{A}\frac{p_{i}^2}{2m A}\right)$ according to \cite{gartenhaus57}.
Applying the binomial theorem for $E$ and substituting into Eq. (\ref{r}), the relativistic kinetic energy $\hat{T}$ takes the form,
\begin{eqnarray}
\nonumber\hat{T}&=&\frac{(\sum _{i}^{A}p_{i})^{2}}{2m A}\\
\label{4}  &=&\frac{1}{2m}\sum_{i=1}^{A}p_{i}^2-\frac{2}{m A}\sum_{i<j}^{A}p_{ij}^{2}
 \end{eqnarray}
Where $p_{ij}=\frac{1}{2}(p_{i}-p_{j})$ is the relative momentum of the two nucleons system. By substituting Eq. (\ref{4}) into Eq. (\ref{1}), this leads to the effective nuclear Hamiltonian operator.
 \begin{equation}\label{r1}
\hat{H}=\sum_{i}^{A} c\vec{\alpha_{i}}.\vec{p_{i}}+(\beta_{i}-I)m_{i}c^2-\frac{1}{2m}p^{2}+ \sum_{i < j}^{A}V_{ij}+\frac{2}{m A}\sum_{i<j}^{A}p_{ij}^{2}
\end{equation}
In Hartree-Fock theory, we seek the best single state given by the lowest energy expectation value of this hamiltonian.
\subsection{Variational and Modified Hartree-Fock Wave function }
One able to ensure the antisymmetry of the fermions' wave functions with the aid of Slater Determinant, and Hartree product to have the convenient form in calculating the ground state energy as the following wave function which is suitable for fermions \cite{jaminon80}. So, the wave function of nucleus $\Psi(r)$ becomes
\begin{equation}
\Psi(r)=\frac{1}{\sqrt{A!}}{det\psi_{i}(\vec{r_{i}})}\\
\end{equation}
Where $\psi_{i}$ is the nucleon wave function which can be expanded as,
\begin{equation}\label{psi}
 \psi_{i}(\vec{r_{i}})=\sum_{\alpha} C_{i\alpha} F_{\alpha}(\vec{r_{i}})
\end{equation}
Where $C_{i\alpha}$ is the oscillator constant, and $F_{\alpha}(\vec{r_{i}})$ is the oscillator wave function which has two components, radial component $\Phi_{\alpha}$ and spin component $\chi_{\alpha}$.
\begin{equation}\label{function}
|F_{\alpha}\rangle =\left|\begin{array}{c}
               \Phi_{\alpha} \\
               \chi_{\alpha}
             \end{array}\right\rangle
 \end{equation}
The two components have the following relation between them\cite{bouyssy87,wen10}, as
\begin{equation}\label{r3}
\chi=\left(1-\frac{\varepsilon-v}{2Mc^2}\right)\frac{\vec{\sigma}.\vec{p}}{2mc}\phi
\end{equation}
The principle of antisymmetry of the wave function did not be honored by the Hartree method according to Slater and Fock independently so, the accurate picture in calculating the ground state energy is the Hartree-Fock approximation. Using Eq. (\ref{r3}), where $\varepsilon$ is external energy, $v$ is a potential energy, and $\vec{\sigma}$ is the Pauli matrices. Here we are dealing with the ground state so, $(\varepsilon-v)/c^{2}$ makes the value of the second term very small and can be neglected.
\begin{equation}\label{roby}
\chi\cong\frac{\vec{\sigma}.\vec{p}}{2mc}\phi
\end{equation}
The wave functions for two nucleons $i$ and $j$ have the formula for bra part $\langle \phi_{\alpha}(r_{i})\phi_{\gamma}(r_{j})|$ and ket part $| \phi_{\beta}(r_{i})\phi_{\delta}(r_{j})\rangle$ as the bracket need two wave functions in each side of bracket,
\begin{eqnarray}
\nonumber\langle \phi_{\alpha}(r_{i})\phi_{\gamma}(r_{j})|&=&\sum_{m_{l_{\alpha}}m_{s_{\alpha}}}\sum_{m_{l_{\gamma}}m_{s_{\gamma}}}
   (l_{\alpha} s_{\alpha} m_{l_{\alpha}} m_{s_{\alpha}}|j_{\alpha} M_{\alpha})(l_{\gamma} s_{\gamma} m_{l_{\gamma}} m_{s_{\gamma}}|j_{\gamma} M_{\gamma})
   \langle\phi_{n_{\alpha}l_{\alpha}m_{l_{\alpha}}}(r_{i})\phi_{n_{\gamma}l_{\gamma}m_{l_{\gamma}}}(r_{j})|\\
   &&\langle\chi_{m_{s_{\alpha}}}^{1/2}
   \chi_{m_{s_{\gamma}}}^{1/2}|\langle\hat{P}_{T_{\alpha}}\hat{P}_{T_{\gamma}}|
\end{eqnarray}
Where $ (l_{\alpha} s_{\alpha} m_{l_{\alpha}} m_{s_{\alpha}}|j_{\alpha} M_{\alpha})$ is the Clebsch-Gordon coefficient, $\chi_{m_{s_{\alpha}}}^{1/2}$is the spin function, and $\hat{P}_{T_{\alpha}}$ is the function of isotopic spin. The two wave functions depend on $r_{i}$ and $r_{j}$ which can be merged to one wave by changing the special coordinates for it, that converts to relative and center of mass coordinates, see \textbf{Appendix A} for more details. Then we have the formula,
\begin{eqnarray}
\nonumber\langle \phi_{\alpha}(r_{i})\phi_{\gamma}(r_{j})|&=&\sum_{m_{l_{\alpha}}m_{s_{\alpha}}}\sum_{m_{l_{\gamma}}m_{s_{\gamma}}}
   \sum_{l S} \sum_{\lambda\mu}\sum_{n l N L}\sum_{m M}\sum_{s m_{s}}\sum_{T}(l_{\alpha} s_{\alpha} m_{l_{\alpha}} m_{s_{\alpha}}|
                                                      j_{\alpha} M_{\alpha})(l_{\gamma} s_{\gamma} m_{l_{\gamma}} m_{s_{\gamma}}|j_{\gamma} M_{\gamma})\\
  \nonumber        &&(l_{\alpha}l_{\gamma}m_{l_{\alpha}}m_{l_{\gamma}}|\lambda\mu)
\langle n_{\alpha}l_{\alpha}n_{\gamma}l_{\gamma}|NL nl\rangle(l S m_{l} m_{S}|J M)(L l M m|\lambda\mu)
 (s_{\alpha}s_{\gamma}m_{s_{\alpha}}m_{s_{\gamma}}|S M_{s})\\
\label{234} &&(s_{\alpha}s_{\gamma}T_{\alpha}T_{\gamma}|T M_{T})\langle\phi_{nlm}(r)\phi_{NLM}(R)|\langle\chi_{m_{s}}^{S}(i,j)|\langle\hat{P}_{T}(i,j)|
\end{eqnarray}
Where $\langle n_{\alpha}l_{\alpha}n_{\gamma}l_{\gamma}|NL nl\rangle$ is the Talmi-Moshinsky bracket and $\phi_{nlm}(r)=R_{nlm}Y_{nlm}$, with radial function $R_{nlm}$ and $Y_{nlm}$ the spherical harmonics, the same treatment happens to the ket part. The bracket of spin function is $\langle\chi_{m_{s}}^{S}(i,j)|\chi_{m_{s}}^{S}(i,j)\rangle=1$ and the isotopic function is $\langle\hat{P}_{T}(i,j)|\hat{P}_{T}(i,j)\rangle=1$. This formula is convenient for two body interaction as in Deuteron and the number of nucleons of Helium nucleus should be emerged in equation through adding $\sum_{i<j=1}^{4}$. The bracket for spherical functions equal one as $\vartheta,\varphi$ are not affected here, but the distance $r$ do. We have the solution of Radial wave function as an oscillator with the Laguerre function (Leigh, Ritz and Galerkin) method \cite{stakgold} where the wave function can be expanded in term of a complete set with basis set
\begin{equation}\label{r6}
R_{nlm}=\left[\frac{2n!}{\Gamma(n+l+\frac{3}{2})}\right]^{\frac{1}{2}}\left(\frac{1}{b}\right)^{\frac{3}{2}}\left(\frac{r}{b}\right)^{l}
exp\left(\frac{-1}{2}\left(\frac{r}{b}\right)^{2}\right)L_{n}^{l+\frac{1}{2}}\left(\frac{r}{b}\right)^{2}
\end{equation}	
$l$ represents the angular momentum, $L_{n}^{l+\frac{1}{2}}$ is the associated Laguerre polynomial\cite{lebeddev}, and the length parameter $b=\sqrt{\frac{(\hbar c)^{2}}{m c^{2}\; \hbar \omega}}$ where $m$ is the mass of the considered particles (nucleons) and $\omega$ is the oscillator frequency. The simplest shell-model should have the overall size of nucleus through the scale of this parameter and it is related to the number density of nucleons or equivalently for $\hbar \omega= 45 A^{-\frac{1}{3}}-25 A^{-\frac{2}{3}}$ according to the equilibrium density $A$ of the even-even nucleus \cite{Kirson}.

\subsection{The handling of the kinetic energy term}
 \par
Using Eqs. (\ref{psi}), (\ref{function}), and (\ref{r1}), we obtain the relativistic modified Hartree-Fock equations, we apply the Lagrange multiplier method for seeking the minimum point of the expression,
\begin{eqnarray}\label{r2}
\nonumber\sum_{i\alpha\beta}h_{i}C_{i\alpha}^{*}C_{i\beta}\left\langle F_{\alpha}|\widetilde{F_{\beta}}\right\rangle&=&\sum_{i\alpha\beta}C_{i\alpha}^{*}C_{i\beta}\left\langle F_{\alpha}(r)|c \vec{\alpha_{i}}.\vec{p_{i}}+(\beta_{i}-I)m_{i}c^2-\frac{1}{2m}p_{i}^{2}|F_{\beta}\right\rangle \\
 \label{rabab2}&+&\sum_{i<j}\sum_{\alpha\gamma\beta\delta}C_{i\alpha}^{*}C_{i\beta}C_{j\gamma}^{*}C_{j\delta} \left\langle F_{\alpha}F_{\gamma}|\frac{2}{A m}p_{ij}^{2}+V_{ij}|\widetilde{F_{\beta}F_{\delta}}\right\rangle
\end{eqnarray}
Differentiate Eq. (\ref{rabab2}) with respect to $C_{i\alpha}^{*}$ which is the conjugate of oscillator constant, one has
\begin{eqnarray}\label{rob3}
\nonumber&&\sum_{i\alpha\beta}C_{i\beta}\left\langle F_{\alpha}|c\alpha_{i}.p_{i}+(\beta_{i}-I)m_{i}c^2-\frac{1}{2m}p_{i}^{2}
|F_{\beta}\right\rangle
+ \sum_{i<j}\sum_{\alpha\gamma\beta\delta}C_{i\beta}C_{j\gamma}^{*}C_{j\delta} \left\langle F_{\alpha}F_{\gamma}|\frac{2}{A m}p_{ij}^{2}+V_{ij}|\widetilde{F_{\beta}F_{\delta}}\right\rangle\\
\label{rabab3}&-&\sum_{i\alpha\beta}h_{i}C_{i\beta}\left\langle F_{\alpha}|F_{\beta}\right\rangle
 = 0
 \end{eqnarray}
Treating the first bracket as $H_{1}$
\begin{equation}
\sum_{i\alpha\beta}C_{i\beta}\left\langle F_{\alpha}|\hat{H_{1}}|F_{\beta}\right\rangle=\sum_{i\alpha\beta}C_{i\beta}\left\langle F_{\alpha}|c\vec{\alpha_{i}}.\vec{p_{i}}+(\beta_{i}-I)m_{i}c^2-\frac{1}{2m}p_{i}^{2}|F_{\beta}\right\rangle
\end{equation}
Taking into account Dirac matrices\cite{Kenneth,green67},
$\alpha=\left(
                        \begin{array}{cc}
                          0 & \sigma \\
                          \sigma & 0 \\
                        \end{array}
                      \right)$
, $\beta=\left(
   \begin{array}{cc}
     I & 0 \\
     0 & -I \\
   \end{array}
 \right)$
 , the unit matrix $ I=\left(
\begin{array}{cc}
 1 & 0 \\
  0 & 1 \\
   \end{array}
   \right) $.
and $p_{i}= p$
    \begin{eqnarray}
\nonumber\left\langle F_{\alpha}| H_{1}|F_{\beta}\right\rangle &=&\left\langle\phi_{\alpha}\left|\frac{(\vec{\sigma}.\vec{p})(\vec{\sigma}.\vec{p})}{2 m }\right|\phi_{\beta}\right\rangle + \left\langle\phi_{\alpha}\left|\frac{(\vec{\sigma}.\vec{p})(\vec{\sigma}.\vec{p})}{2 m }\right|\phi_{\beta}\right\rangle
-\left\langle\phi_{\alpha}\left|\frac{(\vec{\sigma}.\vec{p})(\vec{\sigma}.\vec{p})}{2 m }\right|\phi_{\beta}\right\rangle\\
\nonumber &-&\left\langle\phi_{\alpha}\left|\frac{(p^{2})}{2 m}\right|\phi_{\beta}\right\rangle-\left\langle\phi_{\alpha}\left|\frac{(\vec{\sigma}.\vec{p})(\vec{\sigma}.\vec{p})p^{2}} {(2 m)(4 m^{2}c^{2})}\right|\phi_{\beta}\right\rangle\\
&=& 0
\end{eqnarray}
We ignore the last term for both simplicity and avoiding the fourth power of momentum and speed of light $(\frac{p^{4}}{8m^{3}c^{4}})$, hence we have the kinetic term being vanished.
\subsection{The construction of the potential through One Boson Exchange}
\par
After the treatment of the kinetic energy, the two-body Hamiltonian becomes
\begin{eqnarray}
\label{ground} H &=&\sum_{i<j}(\frac{2}{A m}P_{ij}^{2}+V_{ij})
\end{eqnarray}
The first term in right hand side is the remainder part results from the treatment of kinetic term in Dirac equation as Eq. (\ref{4}) and $V_{ij}(r)$ is the potential according to two-body interaction. The relativistic form of one meson exchange potential between two nucleons $(i,j)$ based on the degrees of freedom associated with three, pseudoscalar, scalar and vector mesons
\begin{equation}\label{mohamed}
 V_{ij}(r)= V_{\pi}(r)+V_{\sigma}(r)+ V_{\omega}(r)
\end{equation}
\begin{eqnarray}\label{mohamed2}
V_{ps}(r) =\gamma_{i}^{o} \gamma_{i}^{5}\gamma_{j}^{o}\gamma_{5}^{j}J_{ps}\;\;,\;\;
 V_{\sigma}=-\gamma_{i}^{o}\gamma_{j}^{o}J_{\sigma}\;\;\;and\;\;\;\; V_{\omega}(r)=\gamma_{i}^{o}\gamma_{j}^{o}
 \overrightarrow{\gamma_{i}}^{\mu}\vec{\gamma_{j}}^{\mu}J_{\omega}\;\;\;where\;\;\;\; \vec{\gamma_{i}}^{\mu}\vec{\gamma_{j}}^{\mu}=[\gamma_{i}^{o}\gamma_{j}^{o}-\vec{\gamma_{i}}\vec{\gamma_{j}}].
\end{eqnarray}
\begin{eqnarray}\label{mohamed3}
 \gamma_{i}^{o}=\left(
                  \begin{array}{cc}
                    1 & 0 \\
                    0 & -1 \\
                  \end{array}
                \right)\;\;\;\;
 \gamma^{5} = \imath\gamma^{o}\gamma^{1}\gamma^{2}\gamma^{3} =\left(
                          \begin{array}{cc}
                            0 & I \\
                            I & 0 \\
                          \end{array}
                        \right)\;\;\;
 \vec{\gamma_{i}} = \left(\begin{array}{cc}
                   I & 0 \\
                   0 & -I \\
                 \end{array}\right)\left(
                                     \begin{array}{cc}
                                       0 & \sigma_{i} \\
                                       -\sigma_{i} & 0 \\
                                     \end{array}
                                   \right)
 \end{eqnarray}
The Dirac representation for mesons' functions  will be used to have Dirac matrices corresponding to Pauli spin matrices\cite{miller72,anselm99}. Substituting Eq. (\ref{mohamed3}) and Eq. (\ref{mohamed2}) into Eq. (\ref{mohamed}) to get $V_{\pi}$, $V_{\sigma}$, and $V_{\omega}$. We add $\pi$-meson as a pseudoscalar one to the previous two mesons because it ties the mesons with the nucleus as it is the fare one. We seek for the stability of nucleus and the exchange of pion meson increases the stability of the nucleus. The attractive behavior is represented in scalar $(\sigma)$ meson and the repulsive behavior is represented in vector $(\omega)$ meson. So, the physics of nucleon potential is maintained. The Fock exchange between the two wave functions spatially is introduced after interaction in the potential, briefly in the ( $\widetilde{}$ ) symbol on the right part (ket part).

\begin{eqnarray}
\nonumber \langle F_{\alpha}F_{\gamma}|V_{ij}|\widetilde{F_{\beta}F_{\delta}}\rangle&=&\langle F_{\alpha}F_{\gamma}|V_{\pi}(r)|\widetilde{F_{\beta}F_{\delta}}\rangle+ \langle F_{\alpha}F_{\gamma}|V_{\sigma}(r)|\widetilde{F_{\beta}F_{\delta}}\rangle+\langle F_{\alpha}F_{\gamma}|V_{\omega(r)}|\widetilde{F_{\beta}F_{\delta}}\rangle\\
 \nonumber  &=& {\langle\phi_{\alpha}|\langle\phi_{\gamma}|J_{\pi}|\widetilde{\chi_{\beta}\rangle|\chi_{\delta}\rangle}}
                  +{ \langle\phi_{\alpha}|\langle\chi_{\gamma}|J_{\pi}|\widetilde{-\phi_{\beta}\rangle|\chi_{\delta}\rangle}}\\
 \nonumber       &&+ {\langle\chi_{\alpha}|\langle\phi_{\gamma}|J_{\pi}|\widetilde{\chi_{\beta}\rangle|-\phi_{\delta}\rangle}}
                  + {\langle\chi_{\alpha}|\langle\chi_{\gamma}|J_{\pi}|\widetilde{-\phi_{\beta}\rangle|-\phi_{\delta}\rangle}}\\
\nonumber &-&\left\langle\left(
                       \begin{array}{cc}
                         \phi_{\alpha} & \phi_{\gamma} \\
                       \end{array}
                     \right)
  |J_{\sigma}|\widetilde{\left(
                    \begin{array}{cc}
                       \phi_{\beta} & \phi_{\delta} \\
                       \end{array}
                       \right)}\right\rangle +\left\langle\left(
                       \begin{array}{cc}
                         \phi_{\alpha} & \chi_{\gamma} \\
                       \end{array}
                     \right)
  |J_{\sigma}|\widetilde{\left(
                    \begin{array}{cc}
                       \chi_{\beta} & \phi_{\delta} \\
                       \end{array}
                       \right)}\right\rangle\\
 \nonumber&+&\left\langle\left(
                       \begin{array}{cc}
                         \chi_{\alpha} & \phi_{\gamma} \\
                       \end{array}
                     \right)
  |J_{\sigma}|\widetilde{\left(
                    \begin{array}{cc}
                       \phi_{\beta} & \chi_{\delta} \\
                       \end{array}
                       \right)}\right\rangle -\left\langle\left(
                       \begin{array}{cc}
                         \chi_{\alpha} & \chi_{\gamma} \\
                       \end{array}
                     \right)
  |J_{\sigma}|\widetilde{\left(
                    \begin{array}{cc}
                       \chi_{\beta} & \chi_{\delta} \\
                       \end{array}
                       \right)}\right\rangle\\
\nonumber   &+& \left\langle\left(
                       \begin{array}{cc}
                         \phi_{\alpha} & \phi_{\gamma} \\
                       \end{array}
                     \right)
  |J_{\omega}|\widetilde{\left(
                    \begin{array}{cc}
                       \phi_{\beta} & \phi_{\delta} \\
                       \end{array}
                       \right)}\right\rangle +\left\langle\left(
                       \begin{array}{cc}
                         \phi_{\alpha} & \chi_{\gamma} \\
                       \end{array}
                     \right)
  |J_{\omega}|\widetilde{\left(
                    \begin{array}{cc}
                       \chi_{\beta} & \phi_{\delta} \\
                       \end{array}
                       \right)}\right\rangle\\
 \nonumber&+&\left\langle\left(
                       \begin{array}{cc}
                         \chi_{\alpha} & \phi_{\gamma} \\
                       \end{array}
                     \right)
  |J_{\omega}|\widetilde{\left(
                    \begin{array}{cc}
                       \phi_{\beta} & \chi_{\delta} \\
                       \end{array}
                       \right)}\right\rangle +\left\langle\left(
                       \begin{array}{cc}
                         \chi_{\alpha} & \chi_{\gamma} \\
                       \end{array}
                     \right)
  |J_{\omega}|\widetilde{\left(
                    \begin{array}{cc}
                       \chi_{\beta} & \chi_{\delta} \\
                       \end{array}
                       \right)}\right\rangle\\
 \nonumber   &-& \left\langle\left(
                       \begin{array}{cc}
                         \phi_{\alpha} & \phi_{\gamma} \\
                       \end{array}
                     \right)
  |J_{\omega}(\vec{\sigma_{i}}.\vec{\sigma_{j}})|\widetilde{\left(
                    \begin{array}{cc}
                       \chi_{\beta} & \chi_{\delta} \\
                       \end{array}
                       \right)}\right\rangle -\left\langle\left(
                       \begin{array}{cc}
                         \phi_{\alpha} & \chi_{\gamma} \\
                       \end{array}
                     \right)
  |J_{\omega}(\vec{\sigma_{i}}.\vec{\sigma_{j}})|\widetilde{\left(
                    \begin{array}{cc}
                       \phi_{\beta} & \chi_{\delta} \\
                       \end{array}
                       \right)}\right\rangle\\
\label{13} &-&\left\langle\left(
                       \begin{array}{cc}
                         \chi_{\alpha} & \phi_{\gamma} \\
                       \end{array}
                     \right)
  |J_{\omega}(\vec{\sigma_{i}}.\vec{\sigma_{j}})|\widetilde{\left(
                    \begin{array}{cc}
                       \chi_{\beta} & \phi_{\delta} \\
                       \end{array}
                       \right)}\right\rangle -\left\langle\left(
                       \begin{array}{cc}
                         \chi_{\alpha} & \chi_{\gamma} \\
                       \end{array}
                     \right)
  |J_{\omega}(\vec{\sigma_{i}}.\vec{\sigma_{j}})|\widetilde{\left(
                    \begin{array}{cc}
                       \phi_{\beta} & \phi_{\delta} \\
                       \end{array}
                       \right)}\right\rangle
\end{eqnarray}
According to the relation between $\phi$ ,$\chi$ in Eq. (\ref{roby}) and defining the momentum for each nucleon(i,j)\cite{jaminon80,franz92} $ \\ p_{i}=p_{r}+\frac{1}{2}p_{R}$,$p_{j}=-p_{r}+\frac{1}{2}p_{R}$ and $p_{i}=\acute{p_{i}}$, $p_{j}=\acute{p_{j}}$, $p_{r}=p$ \\
Substituting those relations into Eq. (\ref{13}),We will apply some important relations \cite{Jacques2000} on this equation.
We have used two static functions for meson degree of freedom in $NN$ interaction, $(GY)$ and $(SPED)$ with $(k=\pi,\sigma,\omega)$. These forms were used to carry out our calculations for Hartree-Fock problem $(HF)$. The first function \cite{miller72} is represented by
\begin{equation}\label{robyy}
    (J_{k})_{GY}=g_{k}\hbar c \left(\frac{\exp(-\mu_{k}r)}{r}-\frac{\exp(-\lambda_{k}r)}{r}\left(1+\frac{\lambda_{k}^{2}-\mu_{k}^{2}}{2\lambda_{k}}r\right)\right)
\end{equation}
 We have $g_{k}^{2}$ is the meson-nucleon coupling constant, $\lambda_{k}$ is a parameter related to the structure function of the form factor and $\mu_{k}=\frac{mc}{\hbar}$ is the range of i-meson associated with the meson mass. The second function has the form \cite{jaminon80},
\begin{equation}\label{robb}
    (J_{k})_{SPED}=g_{k}\hbar c \left(\frac{\lambda_{k}^{2}}{\lambda_{k}^{2}-\mu_{k}^{2}}\right)\left(\frac{\exp(-\mu_{k}r)}{r}-\frac{\exp(-\lambda_{k}r)}{r}\right)
\end{equation}
The details to obtain the following equation is explained in \textbf{Appendices B, C},
\begin{eqnarray}
\nonumber  \langle F_{\alpha}F_{\gamma}|V_{ij}|\widetilde{F_{\beta}F_{\delta}}\rangle &=& \langle\phi_{\alpha}\phi_{\gamma}|-J_{\sigma}+J_{\omega}+\frac{1}{4 m^{2}c^{2}}[2J_{\sigma}p^{2}-
              2\hbar^{2}\left\{\frac{d J_{\sigma}}{d r}\frac{d}{d r}\right\}+\frac{2}{r}\frac{d J_{\sigma}}{d r}[\vec{S}.\vec{L}]\\
  \nonumber            &+& 2J_{\omega}p^{2}-2\hbar^{2}\left\{\frac{d J_{\omega}}{d r}\frac{d}{d r}\right\}+\frac{2}{r}\frac{d J_{\omega}}{d r}
              [S.L]- 6J_{\omega}p^{2}+6\hbar^{2}\left\{\frac{d J_{\omega}}{d r}\frac{d}{d r}\right\}\\
   \nonumber           &-&\frac{6}{r}\frac{d J_{\omega}}{d r}[\vec{S}.\vec{L}]
              + J_{\omega}(\vec{\sigma_{i}}.\vec{\sigma_{j}})\frac{2}{\hbar^{2}}(S.p)^{2}- J_{\omega}(\vec{\sigma_{i}}.\vec{\sigma_{j}})p^{2}\\
    \nonumber          &+& \frac{2}{\hbar^{2}}(S.p)^{2}J_{\omega}(\vec{\sigma_{i}}.\vec{\sigma_{j}})-p^{2}J_{\omega}(\vec{\sigma_{i}}.\vec{\sigma_{j}})]\\
    \nonumber          &+& \frac{1}{4 m^{2}c^{2}}\left[-J_{\omega}(\vec{\sigma_{i}}.\vec{\sigma_{j}})\frac{2}{\hbar^{2}}(S.P_{R})^{2}+J_{\omega}(\vec{\sigma_{i}}.\vec{\sigma_{j}})
    p_{R}^{2}+(1/2)p_{R}^{2}J_{\sigma}+(1/2)p_{R}^{2}J_{\omega}\right]\\
    \nonumber          &+&\frac{- \hbar^{2}}{m^{2}c^{2}}(2S(S+1)-3)(\frac{d J_{\pi}}{d r}\frac{d}{d r})-\frac{1}{ m c^{2}}(\frac{2}{\hbar^{2}}(S.\hat{n})^{2}-1)J_{\pi}
              (\hbar\omega(2n+l+\frac{3}{2})\\
              &-&\frac{1}{2} m\omega^{2}r^{2})|\widetilde{\phi_{\beta}\phi_{\delta}}\rangle.
\end{eqnarray}
With total spin operator $S$  and the meson function $J(r)$, to simplify the solution and get the result, we suppose the nucleons of equal masses so,
 the relative mass $\mu=\frac{m_{1}m_{2}}{m_{1}+m_{2}}$ and center mass $M=m_{1}+m_{2}$.
 \begin{eqnarray}
\nonumber \langle F_{\alpha}F_{\gamma}|V_{ij}|F_{\beta}F_{\delta}\rangle &=& \langle\phi_{\alpha}\phi_{\gamma}|-J_{\sigma}+J_{\omega}+\frac{1}{8 \mu^{2} c^{2}}\left[-\hbar^{2}\left\{\frac{d J_{\sigma}}{d r}\frac{d}{d r}\right\}
+\frac{1}{r}\frac{d J_{\sigma}}{d r}[\frac{\hbar^{2}}{2}[J(J+1)-L(L+1)-S(S+1)]\right]\\
\nonumber &+& 2\hbar^{2}\left\{\frac{d J_{\omega}}{d r}\frac{d}{d r}\right\}-\frac{2}{r}\frac{d J_{\omega}}{d r}
              \left[\frac{\hbar^{2}}{2}[J(J+1)-L(L+1)-S(S+1)]\right]\\
\nonumber &+&\frac{1}{4 \mu c^{2}}[J_{\sigma}(r)\left((\hbar\omega(2n+l+3/2))-1/2\mu\omega^{2}r^{2}\right)
              -2J_{\omega}(r)\left((\hbar\omega(2n+l+3/2))-1/2\mu\omega^{2}r^{2}\right)\\
\nonumber &+& 4(2S(S+1)-3)(\frac{2}{\hbar^{2}}(S.\hat{n})^{2}-1)J_{\omega}
              \left((\hbar\omega(2n+l+3/2))-1/2\mu\omega^{2}r^{2}\right)]\\
\nonumber   &+&\frac{1}{Mc^2}[- 2(2S(S+1)-3)(\frac{2}{\hbar^{2}}(S.\hat{n})^{2}-1)J_{\omega}\left((\hbar\omega(2N+L+3/2))-(1/2)M\omega^{2}R^{2}\right)\\
\nonumber    & +& (J_{\sigma}+J_{\omega})\left((\hbar\omega(2N+L+3/2))-(1/2)M\omega^{2}R^{2}\right)
              J_{\omega}]\\
 \nonumber     &+&\frac{- \hbar^{2}}{m^{2}c^{2}}(2S(S+1)-3)(\frac{d J_{\pi}}{d r}\frac{d}{d r})-\frac{1}{ m c^{2}}(\frac{2}{\hbar^{2}}(S.\hat{n})^{2}-1)J_{\pi}
              (\hbar\omega(2n+l+\frac{3}{2})\\
\label{potential}              &-&\frac{1}{2} m\omega^{2}r^{2})|\widetilde{\phi_{\beta}\phi_{\delta}}\rangle.
\end{eqnarray}

We substitute $(S.\hat{n})$ from \cite{Varshalovich88}. The determination of the energy eigen values requires the diagonalization of the Hamiltonian matrix whose elements are calculated with the functions of Eq. (\ref{r1}). We have Eq. (\ref{potential}) to show that our model can determine a satisfied results for S-state with the meson functions $J_{\sigma}, J_{\omega}, J_{omega}$, the value of mesons' wave functions depends on distance $r$ which is determined as following:
\begin{itemize}
  \item In case of repulsive meson-exchange $\omega$, the lower value of (r) is taken as the hadron radius $\simeq 0.5 fm$ and the upper value is calculated using the following equation with $R$ is the range of meson and $\mu$ is the mass of meson.
      \begin{equation}\label{range}
        R=\frac{1}{\mu}
      \end{equation}
  \item In case of the attractive meson-exchange $\pi, \sigma$, the upper limit of the previous case is taken as the lower limit in this case and the upper limit is determined using Eq. (\ref{range}).
\end{itemize}
In the present work, we have applied our model to calculate the ground states of $^{2}H$ and $^{4}He$ nuclei (A=2,A=4) respectively. We have determined for two nucleons and four nucleons in ($1S_{\frac{1}{2}}$-state) according $^{n}X_{j}$ where $n=1$, $j=l+s$ and $X$ represents the state. Table(1) is the group of parameters used for $(\pi, \sigma, \omega)$mesons. The set of parameters are I,II that include mass $\mu$, coupling constant $(g)$ and the cut-off $\lambda$ parameters.

\begin{center}
Table $1$. The meson parameters for OBEP for different sets.
\begin{tabular}{ | l| l | l | l |  l | l | }
 \hline       &Ref           & meson        & mass MeV     &coupling constant $g_{i}$      & Cut off parameter $\lambda$\;MeV        \\
 \hline set I &\cite{abdo}&\;\;\; $\pi$    & \;\;\; $138.03$ &\;\;\;\;\;\;\;\;\; $14.9$     &\;\;\;\;\;\;\;\;\;\;\;\;\;\; $2000$      \\
              &              &\;\;\; $\sigma$ &\;\;\; $700$  &\;\;\;\;\;\;\;\;\; $16.07$      &\;\;\;\;\;\;\;\;\;\;\;\;\;\; $2000$      \\
              &              &\;\;\; $\omega$ &\;\;\; $782.6$&\;\;\;\;\;\;\;\;\; $28$      &\;\;\;\;\;\;\;\;\;\;\;\;\;\; $1300$      \\
 \hline set II&\cite{abdo}   &\;\;\; $\pi$    &\;\; $138.03$ &\;\;\;\;\;\;\;\;\; $14.40$     &\;\;\;\;\;\;\;\;\;\;\;\;\;\; $1700$      \\
              &              &\;\;\; $\sigma$ &\;\;\; $710$  &\;\;\;\;\;\;\;\;\; $18.37$      &\;\;\;\;\;\;\;\;\;\;\;\;\;\; $2000$      \\
              &              &\;\;\; $\omega$ &\;\; $782.6$  &\;\;\;\;\;\;\;\;\; $24.50$     &\;\;\;\;\;\;\;\;\;\;\;\;\;\; $1850$\\\hline
                         \end{tabular}
\end{center}
We have determined the ratio $R$, to ensure the accuracy between the calculated results and the experimental data \cite{Tyren66}.
\begin{equation}\label{ads}
    R= \frac{E_{theor.} }{E_{exp.}}
\end{equation}
Where $E_{theor}$ is the calculated ground-state and $E_{exp}$ the experimental one.{\color{blue} We can also determine the binding energy per nucleon $\frac{E}{A}$ for the studied nuclei as \cite{gad2011},
\begin{equation}\label{binding}
    \frac{E}{A}=-\frac{E_{g.s.}}{A}
\end{equation}
with the mass number $A$, and the total ground state energy $E_{g.s.}$}
\section{Results and Discussion }
Table(1) represents the group of parameters used for $(\pi, \sigma, \omega)$mesons. The set of parameters are I,II that include mass of meson, the coupling constant $(g)$ and the cut-off parameter $(\lambda)$. The potential is elaborated to calculate the ground state energies for the $H^{2}$, $He^{4}$ nuclei. The results are listed in tables $(2, 3)$ in comparison with the experimental data. The ratio between the present work and experimental one is estimated for both
cases, in other words by using the potential extracted from GY and SPED functions.

We have examined the potential Eq (\ref{potential}) to calculate the ground state energy of $H^{2}$ and $He^{4}$ nuclei using two static meson functions $(GY, SPED)$ with two sets of parameters listed in Table (1) which shows the different sets of the used parameters and for different exchange mesons, $(\sigma, \omega)$ mesons and $(\pi, \sigma, \omega)$ mesons. The potential for different cases is plotted in Figures$(1 - 8)$.\\

Figures$(1 - 8)$ illustrate the potential $V (r)$ in $(MeV)$ versus $r$ in $(fm)$ for $H^{2}$ and $He^{4}$ nuclei for both cases $(GY, SPED)$ and for different sets of parameters $(I, II)$ $(Table(1))$.\\
The potential energy Eq (\ref{potential}) is illustrated in figs.$(1 - 8)$ by two sets of parameter. We have checked them with different meson-exchange function GY and SPED. So, we categorize our results into two groups:- For set I parameter with $(GY, SPED)$ calculated within $(\sigma, \omega)$, and for set II parameter the same above for both nuclei $(H^2, He^4)$.

 All cases are calculated again within $(\sigma, \omega, \pi)$, in other words by adding the third exchange meson "$\pi$" which works attractively at large $r$. Figs. $(1, 2)$ and $(3, 4)$ represent Category "$I, II$" for two meson exchange respectively and Figs. $(5, 6)$ and $(7, 8)$ represent category "$I,II$" for three-meson exchange. Fig. $(1)$ (left panel) shows the potential by using the GY meson function in which the effect of repulsive potential due to $\omega$-meson appears at quite large distance, while the attractive part does not appear clearly as the depth of the potential is very small, the attractive part began with $r\sim 2.0fm$ nears to the diameter of $H^{2}$ nuclei and finished at $r \sim 0.7fm$.

Fig.(1) (right panel) is calculated by using $SPED$ meson function, in this case a significant attractive potential began with $r\sim 1.1fm$ and ended at $r \sim 0.25fm$. Fig. $(2)$ represents the same manner of $H^4$ nuclei with different values of the potential, the transferred point between the attractive and  repulsive parts is similar to one in fig. $(1)$, and the beginning point of the attractive part is controlled by the diameter of nuclei.  \\

\par It can be noticed that, the depth of the attractive potential extracted by $SPED$ function is greater than the one from $GY$ function for both nuclei figs.$(1, 2)$. But it seems that, using set II,  the behavior of two functions is close to each other from the transferred point and the depth, the difference between two nuclei still the values of the attractive potential, and $(\omega-meson)$ is the master here in which the repulsive potential has more higher values and the effect of $(\sigma)$ has been damped see figs.$(3, 4)$. The effect of third meson exchange $(\pi-meson)$ is added as shown in figs.$(5, 6)$ for set I, and figs. $(7, 8)$ for set $II$. Firstly again for set $I$ figs.$(5, 6)$ represent the potential that behaves as the same before for two-meson exchange, in which the attractive part increased very slowly. The depth of the attractive potential increased significantly, $\pi-meson$ flies at more than $r \sim 1.5fm$, and the transferred point has different value from the one in two-meson exchange by using $SPED$ function. On the other hand, for set $II$ as shown in figs.$(7, 8)$ $GY$ and $SPED$  have an improvement in their transferred point and depth than in two-meson exchange, the $SPED$ function still better than $GY$ function.

\begin{center}
Table $2$. The ground state energy of $^{2}H$ with the aid of Table $1$.
\begin{tabular}
{ | l | l |l | l | l | l | l | l | l | l | l | l |}
\hline parameter  & meson  & Present & Present &\; others &\;\;\; exp. &Ratio  &Ratio&{\color{blue}$\frac{E}{A}$}&{\color{blue}$\frac{E}{A}$} & {\color{blue}$\frac{E}{A}$}\\
sets\cite{abdo}  &  -exchange & work$(GY)$ & work$(SPED)$  &   & \cite{exp, kessler, deuter} & $GY$ & $SPED$&{\color{blue}$GY$}&{\color{blue} $SPED$}& {\color{blue}exp.\cite{binding}}\\
\hline \;\;\; I           &\;\;\;\;\;$\sigma$, $\omega$ &\; $-2.916$&\;\;\;\;\;$-2.041$ & {\color{blue}-2.215\cite{haung}}         &           &\;\;1.311&\;\;0.918&1.458&1.0205&   \\
       \;\;\; II                &                       &\; $-3.486$&\;\;\;\;\;$-1.973$& & $-2.224$ &\;\;1.567&\;\;0.887&1.743&0.9865&1.112 \\
       \;\;\; I            &\;\;\;\;\;$\pi$, $\sigma$, $\omega$   &\; $-2.199$&\;\;\;\;\;$-2.248$&-2.220 &  &\;\;0.989&\;\;1.011 &1.0995&1.124& \\
       \;\;\; II         &             &\; $-2.168$&\;\;\;\;\;$-2.204$ & $\pm0.179$\cite{shehab}  &        &\;\;0.975&\;\;0.991 &1.084&1.102& \\ \hline
   \end{tabular}
   \end{center}
\begin{center}
Table $3$. The ground state energy of $^{4}He$ with the aid of Table $1$.
\begin{tabular}
{ | l | l |l | l | l | l | l | l |l | l|l| }
\hline parameter  & meson  & Present & Present &\;\;\cite{kowalski}& exp. & Ratio  & Ratio&{\color{blue}$\frac{E}{A}$}&{\color{blue}$\frac{E}{A}$} & {\color{blue}$\frac{E}{A}$}\\
sets\cite{abdo}  & -exchange & work$(GY)$ & work$(SPED)$ &   & \cite{Tyren66, helium} & $GY$ & $SPED$&{\color{blue}$GY$}&{\color{blue} $SPED$}& {\color{blue}exp.}\\
\hline \;\;\; I      &\;\;\;\;\;\; $\sigma$, $\omega$&\; $-22.372$   &\;\;\;\;\;\;$-20.238$& &              &\;\;\;1.0966&\;\;\;0.992&5.593&5.0595&\\
       \;\;\; II             &                            &\; $-22.751$  &\;\;\;\;\;\;$-21.556$&-21.385 & -20.4$\pm0.3$&\;\;\;1.115&\;\;\;1.057&5.6877&5.389&5.1\\
       \;\;\; I     &\;\;\;\;\;\; $\pi$, $\sigma$, $\omega$&\; $-22.637$&\;\;\;\;\;\;$-20.375$&  &  &\;\;\;1.109&\;\;\;0.999&5.659&5.0937&\\
       \;\;\; II &   &\; $-21.871$&\;\;\;\;\;$-20.337$&   &        &\;\;\;1.072&\;\;\;0.997&5.4677&5.08425&\\\hline
  \end{tabular}
  \end{center}

We observe from \textbf{Table $(2,3)$} two sets of parameters and how many mesons be exchanged between two nucleons, mesons exchanges with two functions and we listed our results for each one for the selected nuclei. Meanwhile, if the ratio tends to unity, the ground energies would be close to the experimental data. The preferable theoretical value of the $^{2}H$ nucleus in case of using two mesons is $SPED$ function for parameters.I, and by using three mesons, we have the value of $SPED$ in the parameter II to be more accurate than others. It is obvious from table(2) and table(3) the ground energy is close to the data in case of SPED function for set I and set II in comparison with the experimental data. The $^{4}He$ nucleus has little different manner, the theoretical value of $SPED$ function for two mesons in parameter I is better than the value of $GY$ function. In case of handling three mesons, $SPED$ function in the parameter II is the best as shown in Table $3$. Using the ratio relation is useful to ensure that our results for $SPED$ function is better than $GY$ function and our attempt to include more two mesons in $OBEP$ analytically is successful in result improvement. The ratio is getting better result for going on more massive nuclei and encouraged for our potential. We concluded that the used model is well-defined and  compatible with the data and even than other models see \cite{Wiringa,Iyad18}. The deuteron ground state energy is quiet little different from the numerical data in \cite{rois,shehab} as a good sign for our constructing potential analytically.{\color{blue} The calculation of binding energy per nucleon serves our idea of being the $OBEP$ with three and four mesons in case of $SPED$ function, and gives satisfied values for Deuteron and Helium nuclei comparing with the experimental one.}

\section{Conclusion}
In the framework of quasi relativistic formulation, the meson exchange potential helps in obtaining a potential with few number of parameters to calculate the ground state for the light nuclei Deuteron and Helium using two $(\sigma,\omega)$ and three $(\pi,\sigma,\omega)$ mesons exchange. In addition, it was shown that a self-consistent treatment of the semi-relativistic nucleon wave function in nuclear state has a great importance in calculations. The difference in masses of $\sigma$ and $\omega$ mesons would not seriously change the main aspect of the concept of relativistic or semi-relativistic interaction, providing an average potential of cancelation of the repulsive meson $(\omega)$ and the attractive meson $(\sigma)$ in conjunction with a weak long range effect $(\pi)$. This work with $OBEP$ in Dirac-Hartree-Fock equation gives a close relationship to other recent approaches, based upon different formalisms which tended to support this direction.
The ground state energies for $^{2}H$ and $^{4}He$ nuclei are successfully determined through this work, and gives us a hope to continue with more massive nuclei. The nuclear properties are being clear in our trail to include more two mesons to describe the NN interaction through our potential. $SPED$ function has an good ability to give us the better shapes of our potential and also better values for energies. We hope that our potential represents a base for NN interaction with different ranges of energies in following search.

\section{Conflict of Interest }
The authors declare that there is no conflict of interests
regarding the publication of this paper.
\appendixa
\par\noindent
The wave functions for two nucleons $i$ and $j$ have a form with Clebsch-Gordon coefficients.
\begin{eqnarray}
\nonumber&&\langle \phi_{\alpha}(r_{i})\phi_{\gamma}(r_{j})|=\\
\nonumber&&\sum_{m_{l_{\alpha}}m_{s_{\alpha}}}\sum_{m_{l_{\gamma}}m_{s_{\gamma}}}
   (l_{\alpha} s_{\alpha} m_{l_{\alpha}} m_{s_{\alpha}}|j_{\alpha} M_{\alpha})(l_{\gamma} s_{\gamma} m_{l_{\gamma}} m_{s_{\gamma}}|j_{\gamma} M_{\gamma})\\
 &&  \langle\phi_{n_{\alpha}l_{\alpha}m_{l_{\alpha}}}(r_{i})\phi_{n_{\gamma}l_{\gamma}m_{l_{\gamma}}}(r_{j})|
   \langle\chi_{m_{s_{\alpha}}}^{1/2}
   \chi_{m_{s_{\gamma}}}^{1/2}|\langle\hat{P}_{T_{\alpha}}\hat{P}_{T_{\gamma}}|
\end{eqnarray}
Where$(l)$ is the orbital angular momentum, $s_{\gamma}$ is the spin, the total angular momentum $j_{\alpha}=l_{\alpha}+ s_{\alpha}$ , $j_{\gamma}=l_{\gamma}+ s_{\gamma}$,
$ M_{\alpha}= m_{l_{\alpha}}+ m_{s_{\alpha}}$ in which $m_{l_{\alpha}}$ is the projection of orbital quantum number ,$m_{s_{\alpha}}$ is the projection of spin quantum number , $M_{\gamma}= m_{l_{\gamma}}+ m_{s_{\gamma}}$ and $\hat{P}_{T_{\alpha}}$ is the function of isotopic spin. The two wave functions are not connected and depend on $r_{i}$,$r_{j}$ so, the two wave functions need to be connected
\begin{eqnarray}
\nonumber&&\langle \phi_{\alpha}(r_{i})\phi_{\gamma}(r_{j})|=\\
\nonumber&&\sum_{m_{l_{\alpha}}m_{s_{\alpha}}}\sum_{m_{l_{\gamma}}m_{s_{\gamma}}}\sum_{\lambda\mu}
                                                      (l_{\alpha} s_{\alpha} m_{l_{\alpha}} m_{s_{\alpha}}|j_{\alpha} M_{\alpha})(l_{\gamma} s_{\gamma} m_{l_{\gamma}} m_{s_{\gamma}}|j_{\gamma} M_{\gamma})\\
                                                      &&(l_{\alpha}l_{\gamma}m_{l_{\alpha}}m_{l_{\gamma}}|\lambda\mu)
    \langle\phi_{n_{\alpha}l_{\alpha}m_{l_{\alpha}}}(r_{i})\phi_{n_{\gamma}l_{\gamma}m_{l_{\gamma}}}(r_{j})|
                                                     \langle\chi_{m_{s_{\alpha}}}^{1/2}\chi_{m_{s_{\gamma}}}^{1/2}|\langle\hat{P}_{T_{\alpha}}
                                                     \hat{P}_{T_{\gamma}}|
\end{eqnarray}
With $\lambda=l_{\alpha}+l_{\gamma}$ and $\mu=m_{l_{\alpha}}+m_{l_{\gamma}}$, we can change the special coordinates for each wave functions to become one wave, that depends on relative mass and center of mass.
\begin{eqnarray}
\nonumber&&\langle \phi_{\alpha}(r_{i})\phi_{\gamma}(r_{j})|=\\
\nonumber&&\sum_{m_{l_{\alpha}}m_{s_{\alpha}}}\sum_{m_{l_{\gamma}}m_{s_{\gamma}}}
    \sum_{\lambda\mu}\sum_{n l N L}(l_{\alpha} s_{\alpha} m_{l_{\alpha}} m_{s_{\alpha}}|
                                                      j_{\alpha} M_{\alpha})(l_{\gamma} s_{\gamma} m_{l_{\gamma}} m_{s_{\gamma}}|j_{\gamma} M_{\gamma})\\
\nonumber&&(l_{\alpha}l_{\gamma}m_{l_{\alpha}}m_{l_{\gamma}}|\lambda\mu)
    \langle n_{\alpha}l_{\alpha}n_{\gamma}l_{\gamma}|NL nl\rangle
    \langle\phi_{{NL} {nl}}(r,R)|\langle\chi_{m_{s_{\alpha}}}^{1/2}\chi_{m_{s_{\gamma}}}^{1/2}|\\
&&\langle\hat{P}_{T_{\alpha}}\hat{P}_{T_{\gamma}}|
\end{eqnarray}
Where $\langle n_{\alpha}l_{\alpha}n_{\gamma}l_{\gamma}|NL nl\rangle$ is the Talmi-Moshinsky bracket , $NL$ is total center of mass , $nl$ is  total relative. The wave function $\phi_{{NL} {nl}}(r,R)$ can be spitted in to the form
\begin{eqnarray}
\nonumber&&\langle \phi_{\alpha}(r_{i})\phi_{\gamma}(r_{j})|=\\
\nonumber&&\sum_{m_{l_{\alpha}}m_{s_{\alpha}}}\sum_{m_{l_{\gamma}}m_{s_{\gamma}}}\sum_{JM}
    \sum_{\lambda\mu}\sum_{n l N L}\sum_{m M}(l_{\alpha} s_{\alpha} m_{l_{\alpha}} m_{s_{\alpha}}|
                                                      j_{\alpha} M_{\alpha})(l_{\gamma} s_{\gamma} m_{l_{\gamma}} m_{s_{\gamma}}|j_{\gamma} M_{\gamma})\\
 \nonumber                                                     &&(l_{\alpha}l_{\gamma}m_{l_{\alpha}}m_{l_{\gamma}}|\lambda\mu)
    \langle n_{\alpha}l_{\alpha}n_{\gamma}l_{\gamma}|NL nl\rangle(l S m_{l} m_{S}|J M)(L l M m|\lambda\mu)\\
 &&   \langle\phi_{NLM}(R)\phi_{nlm}(r)|
    \langle\chi_{m_{s_{\alpha}}}^{1/2}\chi_{m_{s_{\gamma}}}^{1/2}|\langle\hat{P}_{T_{\alpha}}\hat{P}_{T_{\gamma}}|
\end{eqnarray}
As $L$ gives the total orbital quantum number in center of mass , $l$ gives the total orbital quantum number in relative coordinates and $S=s_{i}+s_{j}$ is the total spin.
Relative to the spin functions and isospin functions to be connected ,we have to use them as following.
 \begin{eqnarray}
\nonumber&&\langle \phi_{\alpha}(r_{i})\phi_{\gamma}(r_{j})|=\\
\nonumber&&\sum_{m_{l_{\alpha}}m_{s_{\alpha}}}\sum_{m_{l_{\gamma}}m_{s_{\gamma}}}\sum_{JM}
    \sum_{\lambda\mu}\sum_{n l N L}\sum_{m M}\sum_{s m_{s}}\sum_{T}(l_{\alpha} s_{\alpha} m_{l_{\alpha}} m_{s_{\alpha}}|
    j_{\alpha} M_{\alpha})(l_{\gamma} s_{\gamma} m_{l_{\gamma}} m_{s_{\gamma}}|j_{\gamma} M_{\gamma})\\
\nonumber &&(l_{\alpha}l_{\gamma}m_{l_{\alpha}}m_{l_{\gamma}}|\lambda\mu)
    \langle n_{\alpha}l_{\alpha}n_{\gamma}l_{\gamma}|NL nl\rangle(l S m_{l} m_{S}|J M)(L l M m|\lambda\mu)\\
&& (s_{\alpha}s_{\gamma}m_{s_{\alpha}}m_{s_{\gamma}}|S M_{s})
 (s_{\alpha}s_{\gamma}T_{\alpha}T_{\gamma}|T M_{T})\langle\phi_{NLM}(R)\phi_{nlm}(r)|\langle\chi_{m_{s}}^{S}(i,j)|\langle\hat{P}_{T}(i,j)|
\end{eqnarray}
 With $T_{proton}=\frac{-1}{2}$ and $T_{neutron}=\frac{1}{2}$.
\appendixb
\par\noindent
According to the relation between $\phi$ , $\chi$ in Eq.(\ref{roby}), one obtain
\begin{eqnarray}
\nonumber &&\langle F_{\alpha}F_{\gamma}| V_{ij}(r)|\widetilde{F_{\beta}F_{\delta}}\rangle \\
\nonumber&&=\langle\phi_{\alpha}\phi_{\gamma}|-J_{\sigma}+J_{\omega}+\frac{1}{4 m^{2}c^{2}} [\vec{(\sigma_{i}}.\vec{P_{i}})J_{\sigma}(\vec{\sigma_{i}}.\acute{\vec{p_{i}}})+(\vec{\sigma_{j}}.\vec{P_{j}})J_{\sigma}(\vec{\sigma_{j}}.\acute{\vec{p_{j}}})\\
\nonumber&&+(\vec{\sigma_{i}}.\vec{p_{i}})J_{\omega}(\vec{\sigma_{i}}.\acute{\vec{p_{i}}})+(\vec{\sigma_{j}}.\vec{p_{j}})J_{\omega}(\vec{\sigma_{j}}.\acute{\vec{p_{j}}})
-J_{\omega}(\vec{\sigma_{i}}.\vec{\sigma_{j}})(\vec{\sigma_{j}}.\acute{\vec{p_{j}}})(\vec{\sigma_{i}}.\acute{\vec{p_{i}}})\\
\nonumber&&-(\vec{\sigma_{j}}.\vec{p_{j}})J_{\omega}(\vec{\sigma_{i}}.\vec{\sigma_{j}})(\vec{\sigma_{i}}.\acute{\vec{p_{i}}})-(\vec{\sigma_{i}}.\vec{p_{i}})J_{\omega} (\vec{\sigma_{i}}.\vec{\sigma_{j}})(\vec{\sigma_{j}}.\acute{\vec{p_{j}}})\\
\label{13}&&-(\vec{\sigma_{i}}.\vec{p_{i}})(\vec{\sigma_{j}}.\vec{p_{j}})J_{\omega}(\vec{\sigma_{i}}.\vec{\sigma_{j}})]|\widetilde{\phi_{\beta}\phi_{\delta}}\rangle
 \end{eqnarray}
Defining the momentum for each nucleon (i, j) $p_{i}=p_{R}+\frac{1}{2}p_{r}$, $p_{j}=-p_{r}+\frac{1}{2}p_{r}$ \\and $p_{i}=\acute{p_{i}}$, $p_{j}=\acute{p_{j}}$ \cite{neff15,franz92}. Substituting those relations into Eq. (\ref{13}), the dependence of $J(r)$ on the relative distance $(r)$ not on $(R)$ makes its movement with the center of mass operators more easy,\\
where $p_{r}=p$ and $(\sigma_{i}.p_{r})(\sigma_{i}.p_{r})=p_{r}^{2}$, we obtain
\begin{eqnarray}
\nonumber&&\langle F_{\alpha}F_{\gamma}| V_{ij}(r)|\widetilde{F_{\beta}F_{\delta}}\rangle \\
 \nonumber&&=\langle\phi_{\alpha}\phi_{\gamma}|-J_{\sigma}+J_{\omega}+\frac{1}{4 m^{2}c^{2}}[(\vec{\sigma_{i}}.\vec{p})J_{\sigma}(\vec{\sigma_{i}}.\vec{p})
              + (\vec{\sigma_{j}}.\vec{p})J_{\sigma}(\vec{\sigma_{j}}.\vec{p})\\
              \nonumber&&+(\vec{\sigma_{i}}.\vec{p})J_{\omega}
              (\vec{\sigma_{i}}.\vec{p})+(\vec{\sigma_{j}}.\vec{p})J_{\omega}(\vec{\sigma_{j}}.\vec{p})+ J_{\omega}(\vec{\sigma_{i}}.\vec{\sigma_{j}})(\vec{\sigma_{j}}.\vec{p})(\vec{\sigma_{i}}.\vec{p})\\
       \nonumber&&+(\vec{\sigma_{i}}.\vec{p})J_{\omega}(\vec{\sigma_{i}}.\vec{\sigma_{j}})(\vec{\sigma_{j}}.\vec{p})+(\vec{\sigma_{j}}.\vec{p})
       J_{\omega}(\vec{\sigma_{i}}.\vec{\sigma_{j}})(\vec{\sigma_{i}}.\vec{p})+(\vec{\sigma_{i}}.\vec{p})
              (\vec{\sigma_{j}}.\vec{p})\\
             \nonumber&& J_{\omega}(\vec{\sigma_{i}}.\vec{\sigma_{j}})+ J_{\omega}(\vec{\sigma_{i}}.\vec{\sigma_{j}})(\vec{\sigma_{j}}.\vec{p})(\vec{\vec{\sigma_{i}}}.\vec{p_{R}})- J_{\omega}(\vec{\sigma_{i}}.\vec{\sigma_{j}})(\vec{\sigma_{j}}.\vec{p_{r}})(\vec{\sigma_{i}}.\vec{p})\\
              \nonumber&&+(\vec{\sigma_{j}}.\vec{p})(\vec{\sigma_{i}}.\vec{p_{r}})J_{\omega}(\vec{\sigma_{i}}.\vec{\sigma_{j}})
              -(\vec{\sigma_{i}}.\vec{p})(\vec{\sigma_{j}}.\vec{p_{r}})
              J_{\omega}(\vec{\sigma_{i}}.\vec{\sigma_{j}})]\\
 \nonumber&&+ \frac{1}{8 m^{2}c^{2}}[(\vec{\sigma_{i}}.\vec{p})(\vec{\sigma_{i}}.\vec{p_{r}})J_{\sigma}+J_{\sigma}(\vec{\sigma_{i}}.\vec{p_{r}})(\vec{\sigma_{i}}.\vec{p})
              - (\vec{\sigma_{j}}.\vec{p})(\vec{\sigma_{j}}.\vec{p_{r}})J_{\sigma}\\
              \nonumber&&-J_{\sigma}(\vec{\sigma_{j}}.\vec{p_{r}})
              (\vec{\sigma_{j}}.\vec{p})+(\vec{\sigma_{i}}.\vec{p})(\vec{\sigma_{i}}.\vec{p_{r}})J_{\omega}+ J_{\omega}(\vec{\sigma_{i}}.\vec{p_{r}})(\vec{\sigma_{i}}.\vec{p})\\
              \nonumber&&-(\vec{\sigma_{j}}.\vec{p})(\vec{\sigma_{j}}.\vec{p_{r}})J_{\omega}
              -J_{\omega}(\vec{\sigma_{j}}.\vec{p_{r}})(\vec{\sigma_{j}}.\vec{p})+p_{r}^{2}J_{\sigma}
              +p_{r}^{2}J_{\omega}\\
              &&-2 J_{\omega}(\vec{\sigma_{i}}.\vec{\sigma_{j}})(\vec{\sigma_{j}}.\vec{p_{r}})(\vec{\sigma_{i}}.\vec{p_{r}})]
              |\widetilde{\phi_{\beta}\phi_{\delta}}\rangle
\end{eqnarray}
 We will apply some important relations \cite{Raynal}
\begin{enumerate}
\item $(\vec{\sigma_{1}}.\vec{A})(\vec{\sigma_{1}}.\vec{B})=A.B+\imath\vec{\sigma_{1}}(A\times B)$
\item $(\vec{\sigma_{1}}.\vec{A})^{2}=A^{2}$
\item$(\vec{\sigma_{1}}.\vec{A})(\vec{\sigma_{2}}.\vec{A})=\frac{2}{\hbar^{2}}(S.A)^{2}-A^{2}$
\item $(\vec{\sigma}.\vec{A})F(r)(\vec{\sigma}.\vec{A})=F(r)A^{2}-\imath\hbar\{\bigtriangledown F(r).A+\imath\sigma[(\bigtriangledown F(r))\times A]\}$
\item $\vec{\sigma_{i}}\vec{\sigma_{j}} = 2\delta_{ij}-\sigma_{ji}$
\end{enumerate}
Including these relations in potential equation.
\begin{eqnarray}
\nonumber &&\langle F_{\alpha}F_{\gamma}| V_{ij}(r)|\tilde{F_{\beta}F_{\delta}}\rangle\\
\nonumber&&=\langle\phi_{\alpha}\phi_{\gamma}|-J_{\sigma}+J_{\omega}+\frac{1}{4 m^{2}c^{2}}[(\vec{\sigma_{i}}.\vec{p})J_{\sigma}(\vec{\sigma_{i}}.\vec{p})
                      + (\vec{\sigma_{j}}.\vec{p})J_{\sigma}(\vec{\sigma_{j}}.\vec{p})\\
\nonumber             &&+(\vec{\sigma_{i}}.\vec{p})J_{\omega}(\vec{\sigma_{i}}.\vec{p})+(\vec{\sigma_{j}}.\vec{p})J_{\omega}(\vec{\sigma_{j}}.\vec{p})
             + J_{\omega}(\vec{\sigma_{i}}.\vec{\sigma_{j}})(\vec{\sigma_{j}}.\vec{p})(\vec{\sigma_{i}}.\vec{p})\\
\nonumber   &&+ (\vec{\sigma_{i}}.\vec{p})J_{\omega}(\vec{\sigma_{i}}.\vec{\sigma_{j}})(\vec{\sigma_{j}}.\vec{p})+(\vec{\sigma_{j}}.\vec{p})J_{\omega}
(\vec{\sigma_{i}}.\vec{\sigma_{j}})(\vec{\sigma_{i}}.\vec{p})+ (\vec{\sigma_{i}}.\vec{p})(\vec{\sigma_{j}}.\vec{p})\\
\nonumber&&J_{\omega}(\vec{\sigma_{i}}.\vec{\sigma_{j}})]
+ \frac{1}{4 m^{2}c^{2}}\left[J_{\omega}(\vec{\sigma_{i}}.\vec{\sigma_{j}})(\vec{\sigma_{j}}.\vec{p_{r}})(\vec{\sigma_{i}}.\vec{p_{r}})+
                      (1/2)p_{r}^{2}J_{\sigma}+(1/2)p_{r}^{2}J_{\omega}\right]\\
\label{moh3}&&|\tilde{\phi_{\beta}\phi_{\delta}}\rangle
 \end{eqnarray}
To get the solution of Eq. (\ref{moh3}) we substitute every term as following, using the relation of angular momentum $L=\vec{r}\times p$, $\sigma=\frac{2S}{\hbar}$ where $S$ is the spin operator, $P=-\imath\hbar\nabla$, and $\nabla J_{\sigma}=\frac{1}{r}(\frac{d J_{\sigma}}{d r})r$. According to the previous relations, and where $\sigma_{j}^{2}=\sigma_{x}^{2}+\sigma_{y}^{2}+\sigma_{z}^{2}=1$ as triplet case for two nucleons
 \begin{eqnarray}
\nonumber(\vec{\sigma_{j}}.\vec{p})J_{\omega}(r)(\vec{\sigma_{j}}.\vec{p})&=&(\vec{\sigma_{j}}.\vec{p})J_{\omega}(r)
\sigma_{j}^{2}(\sigma_{i}.p)\\
\nonumber     &=&-3J_{\omega}(r)p^{2}+3\hbar^{2}\left\{\frac{d J_{\omega}}{d r}\frac{d}{d r}\right\}\\
     &-&\frac{6}{r}\frac{d J_{\omega}}{d r}[\vec{S_{j}}.\vec{L}]
\end{eqnarray}
 \begin{eqnarray}
\nonumber   (\vec{\sigma_{i}}.\vec{p})J_{\omega}(r)(\vec{\sigma_{i}}\vec{\sigma_{j}})(\vec{\sigma_{j}}.\vec{p})&=& -3 J_{\omega}(r)p^{2}+3\hbar^{2}\left\{\frac{d J_{\omega}}{d r}\frac{d}{d r}\right\}\\
   &-&\frac{6}{r}\frac{d J_{\omega}}{d r}[\vec{S_{i}}.\vec{L}]
 \end{eqnarray}
We substitute those terms in Eq. (\ref{moh3})
\begin{eqnarray}
\nonumber&&  \langle F_{\alpha}F_{\gamma}| V_{ij}(r)|\tilde{F_{\beta}F_{\delta}}\rangle \\
\nonumber&&= \langle\phi_{\alpha}\phi_{\gamma}|-J_{\sigma}+J_{\omega}+\frac{1}{4 m^{2}c^{2}}[2J_{\sigma}(r)p^{2}-
              2\hbar^{2}\left\{\frac{d J_{\sigma}}{d r}\frac{d}{d r}\right\}+\frac{2}{r}\frac{d J_{\sigma}}{d r}[\vec{S}.\vec{L}]\\
\nonumber     &&- 4J_{\omega}(r)p^{2}+4\hbar^{2}\left\{\frac{d J_{\omega}}{d r}\frac{d}{d r}\right\}-\frac{4}{r}\frac{d J_{\omega}}{d r}
              [S.L]
              + J_{\omega}(\vec{\sigma_{i}}.\vec{\sigma_{j}})\frac{2}{\hbar^{2}}(S.p)^{2}\\
\nonumber      &&- J_{\omega}(\vec{\sigma_{i}}.\vec{\sigma_{j}})p^{2}
              + \frac{2}{\hbar^{2}}(S.p)^{2}J_{\omega}(\vec{\sigma_{i}}.\vec{\sigma_{j}})-p^{2}J_{\omega}(\vec{\sigma_{i}}.\vec{\sigma_{j}})]\\
             \nonumber &&+ \frac{1}{4 m^{2}c^{2}}[-J_{\omega}(\vec{\sigma_{i}}.\vec{\sigma_{j}})\frac{2}{\hbar^{2}}(S.p_{r})^{2}+J_{\omega}(\vec{\sigma_{i}}.\vec{\sigma_{j}})p_{r}^{2}
              +(1/2)p_{r}^{2}J_{\sigma}\\
              &&+(1/2)p_{r}^{2}J_{\omega}]|\tilde{\phi_{\beta}\phi_{\delta}}\rangle
\end{eqnarray}
With total spin operator $S$  and the meson function $J(r)$, using\\
$(S.P)^{2}=(S.\hat{n})^{2}P^{2}$, $(\sigma_{i}.\sigma_{j})=\frac{2}{\hbar^{2}}S^{2}-3$ and $S.L=\frac{\hbar^{2}}{2}[J(J+1)-L(L+1)-S(S+1)]$ \cite{Varshalovich88}.
Quantum mechanics have a magnificent tool, this tool is the harmonic oscillator which is capable  of being solved in closed form, it has generally useful approximations and exact solutions of different problems \cite{Kirson}. It solves the differential equations in quantum mechanics.
We have the energy of Harmonic Oscillator$(\hbar\omega(2n+l+3/2))$ which equals the kinetic energy$(\frac{P^{2}}{2 m})$ added to the potential energy$((1/2)m\omega^{2}x^{2})$ to simplify the solution and get the result. It is slitted in relative harmonic oscillator energy
$\hbar\omega(2n+l+\frac{3}{2}) = \frac{P^{2}}{2 \mu}+\frac{1}{2}\mu\omega^{2}r^{2}$ \cite{gartenhaus57,Gad}, with $\omega$ that is the angular frequency, and center of mass contribution in harmonic oscillator energy $\hbar\omega(2N+L+\frac{3}{2})= \frac{P^{2}}{2 M}+\frac{1}{2}M\omega^{2}R^{2}$.
\newpage
We suppose the nucleons of equal masses, so the relative mass $\mu=\frac{m_{1}m_{2}}{m_{1}+m_{2}}=\frac{m}{2}$ and center mass $M=m_{1}+m_{2}=2m$.
\begin{eqnarray}
\nonumber&& \langle F_{\alpha}F_{\gamma}| V_{ij}(r)|\tilde{F_{\beta}F_{\delta}}\rangle \\
\nonumber&&= \langle\phi_{\alpha}\phi_{\gamma}|-J_{\sigma}+J_{\omega}+\frac{1}{8 \mu^{2} c^{2}}[-\hbar^{2}\{\frac{d J_{\sigma}}{d r}\frac{d}{d r}\}+\frac{1}{r}\frac{d J_{\sigma}}{d r}[\frac{\hbar^{2}}{2}[J(J+1)\\
\nonumber&&-L(L+1)-S(S+1)]]+ 2\hbar^{2}\left\{\frac{d J_{\omega}}{d r}\frac{d}{d r}\right\}-\frac{2}{r}\frac{d J_{\omega}}{d r}
             [\frac{\hbar^{2}}{2}[J(J+1)\\
             \nonumber&&-L(L+1)-S(S+1)]]+\frac{1}{4 \mu c^{2}}[J_{\sigma}(r)(\frac{p^{2}}{2\mu})-
              2J_{\omega}(r)(\frac{p^{2}}{2\mu})+ 2J_{\omega}\\
              \nonumber&&(2S(S+1)-3)(\frac{2}{\hbar^{2}}(S.\hat{n})^{2}-1)(\frac{p^{2}}{2\mu})+2(\frac{2}{\hbar^{2}}(S.\hat{n})^{2}-1)
              (\frac{p^{2}}{2\mu})J_{\omega}\\
              \nonumber&&(2S(S+1)-3)]+\frac{1}{Mc^2}[-2(2S(S+1)-3)J_{\omega}(r)(\frac{2}{\hbar^{2}}(S.\hat{n})^{2}-1)\\
              \nonumber&&(\frac{p_{r}^{2}}{2M})
 + (\frac{p_{r}^{2}}{2M})J_{\sigma}+(\frac{p_{r}^{2}}{2M})
              J_{\omega}]|\tilde{\phi_{\beta}\phi_{\delta}}\rangle
\end{eqnarray}
The total formula of two-nucleons interaction through the exchange of two mesons where $p_{ij}=p$ and $A1$ is the mass number of the required nuclei.
\begin{eqnarray}
\nonumber&& \langle \phi_{\alpha}(r_{i})\phi_{\gamma}(r_{j})|H|\phi_{\beta}(r_{i})\phi_{\delta}(r_{j})\rangle \\
\nonumber&&= \sum_{m_{l_{\alpha}}m_{s_{\alpha}}}\sum_{m_{l_{\gamma}}m_{s_{\gamma}}}\sum_{JM}
    \sum_{\lambda\mu}\sum_{n l N L}\sum_{m M}\sum_{s m_{s}}\sum_{T}(l_{\alpha} s_{\alpha} m_{l_{\alpha}} m_{s_{\alpha}}|j_{\alpha} M_{\alpha})
   (l_{\gamma} s_{\gamma} m_{l_{\gamma}} m_{s_{\gamma}}|j_{\gamma} M_{\gamma})\\
     \nonumber &&(l_{\alpha}l_{\gamma}m_{l_{\alpha}}m_{l_{\gamma}}|\lambda\mu)
  \langle n_{\alpha}l_{\alpha}n_{\gamma}l_{\gamma}|NL nl\rangle(l S m_{l} m_{S}|J M)(L l M m|\lambda\mu)
 (s_{\alpha}s_{\gamma}m_{s_{\alpha}}m_{s_{\gamma}}|S M_{s})\\
\nonumber && (\chi_{\alpha}\chi_{\gamma}T_{\alpha}T_{\gamma}|M_{T} T)
 \langle R_{N LM}(R)Y_{N LM} R_{nlm}(r)Y_{nlm}|\langle\hat{P}_{T}(i,j)|\\
 \nonumber&&\frac{4}{A1}((\hbar\omega(2n+l+3/2))-1/2\mu\omega^{2}r^{2})-J_{\sigma}+J_{\omega}\\
 \nonumber&&+\frac{1}{8 \mu^{2} c^{2}}[-\hbar^{2}\{\frac{d J_{\sigma}}{d r}\frac{d}{d r}\}+\frac{1}{r}\frac{d J_{\sigma}}{d r}[\frac{\hbar^{2}}{2}[J(J+1)-L(L+1)-S(S+1)]]\\
 \nonumber    &&+ 2\hbar^{2}\{\frac{d J_{\omega}}{d r}\frac{d}{d r}\}-\frac{2}{r}\frac{d J_{\omega}}{d r}
             [\frac{\hbar^{2}}{2}[J(J+1)-L(L+1)-S(S+1)]]]\\
   \nonumber &&+\frac{1}{4 \mu c^{2}}[J_{\sigma}(r)((\hbar\omega(2n+l+3/2))-1/2\mu\omega^{2}r^{2})-
              2J_{\omega}(r)((\hbar\omega(2n+l+3/2))\\
              \nonumber&&-1/2\mu\omega^{2}r^{2})+ 2J_{\omega}(2S(S+1)-3)(\frac{2}{\hbar^{2}}(S.\hat{n})^{2}-1)((\hbar\omega(2n+l+3/2))\\
              \nonumber&&-1/2\mu\omega^{2}r^{2})+2(\frac{2}{\hbar^{2}}(S.\hat{n})^{2}-1)((\hbar\omega(2n+l+3/2))-1/2\mu\omega^{2}r^{2})J_{\omega}\\
              \nonumber&&(2S(S+1)-3)]+\frac{1}{Mc^2}[-2(2S(S+1)-3)J_{\omega}(r)(\frac{2}{\hbar^{2}}(S.\hat{n})^{2}-1)\\
              \nonumber&&((\hbar\omega(2N+L+3/2))-(1/2)M\omega^{2}R^{2})+ ((\hbar\omega(2N+L+3/2))\\
              \nonumber&&-(1/2)M\omega^{2}R^{2})J_{\sigma}+((\hbar\omega(2N+L+3/2))-(1/2)M\omega^{2}R^{2})
              J_{\omega}]|\\
\nonumber&& \sum_{m_{l_{\beta}}m_{s_{\beta}}}\sum_{m_{l_{\delta}}m_{s_{\delta}}}\sum_{JM}
   \sum_{\lambda\mu}\sum_{n l N L}\sum_{m M}\sum_{s m_{s}}\sum_{T}(l_{\beta} s_{\beta} m_{l_{\beta}} m_{s_{\beta}}|j_{\beta} M_{\beta})\\
\nonumber && (l_{\delta} s_{\delta} m_{l_{\delta}} m_{s_{\delta}}|j_{\delta} M_{\delta})
(l_{\alpha}l_{\gamma}m_{l_{\alpha}}m_{l_{\gamma}}|\lambda\mu)
 \langle n_{\beta}l_{\beta}n_{\delta}l_{\delta}|NL nl\rangle(l S m_{l} m_{S}|J M)(L l M m|\lambda\mu)\\
\nonumber&& (s_{\beta}s_{\delta}m_{s_{\beta}}m_{s_{\delta}}|S M_{s})
(\chi_{\beta}\chi_{\delta}T_{\beta}T_{\delta}|M_{T} T)
  |R_{NLM}(R)Y_{NLM} R_{nlm}(r)Y_{nlm}\rangle\\
 && |\hat{{P}}_{T}(i,j)\rangle
\end{eqnarray}
\newpage
\appendixc
\par\noindent
 We choose pion $\pi$ as a pseudoscalar meson to be added to the previous two mesons, because it is the one which tie the mesons with nucleus as it is the fare one. We do not choose another pseudoscalar meson as it demands a reaction between two nucleons and we want to calculate the ground state, so we seek for stability of nucleus and the exchange of pion meson increases the stability of nucleus. We have the pseudoscalar potential as
\begin{eqnarray}
\nonumber&& \langle F_{\alpha}F_{\gamma}|V_{ps}(r)|\tilde{F_{\beta}F_{\delta}}\rangle\\
\nonumber&&=\langle\left(
                       \begin{array}{cc}
                         \phi_{\alpha} & \chi_{\alpha} \\
                       \end{array}
                     \right)
  |\langle\left(
            \begin{array}{cc}
              \phi_{\gamma} & \chi_{\gamma} \\
            \end{array}
          \right)
  |\left(\begin{array}{cc}
                1 & 0 \\
                0 & -1
                \end{array}\right)_{i}\left(
                \begin{array}{cc}
                0 & 1 \\
                 1 & 0 \\
                  \end{array}
                  \right)_{i}\left(
                  \begin{array}{cc}
                   1 & 0 \\
                    0 & -1 \\
                     \end{array}
                      \right)_{j}\\
                      \nonumber&&\left(
                                                                                                                                                                              \begin{array}{cc}
                       0 & 1 \\
                       1 & 0 \\
                       \end{array}
                       \right)_{j} J_{\pi}
                        |\tilde{\left(
                       \begin{array}{c}
                       \phi_{\beta} \\
                       \chi_{\beta} \\
                       \end{array}
                       \right)\rangle|
                       \left(
                       \begin{array}{c}
                       \phi_{\delta} \\
                       \chi_{\delta} \\
                       \end{array}
                       \right)}\rangle\\
  \nonumber   &&={\langle\phi_{\alpha}|\langle\phi_{\gamma}|J_{\pi}|\tilde{\chi_{\beta}\rangle|\chi_{\delta}}\rangle}
                  +{ \langle\phi_{\alpha}|\langle\chi_{\gamma}|J_{\pi}|\tilde{-\phi_{\beta}\rangle|\chi_{\delta}\rangle}}\\
        &&+ {\langle\chi_{\alpha}|\langle\phi_{\gamma}|J_{\pi}|\tilde{\chi_{\beta}\rangle|-\phi_{\delta}}\rangle}
                  + {\langle\chi_{\alpha}|\langle\chi_{\gamma}|J_{\pi}|\tilde{-\phi_{\beta}\rangle|-\phi_{\delta}}\rangle}
   \end{eqnarray}
Substituting from the previous relations in the treatment of two mesons, we obtain
 \begin{eqnarray}
\nonumber&& \langle F_{\alpha}F_{\gamma}|V_{ps}(r)|\tilde{F_{\beta}F_{\delta}}\rangle  \\
\nonumber&&=  \langle\left(
                       \begin{array}{cc}
                         \phi_{\alpha} & \phi_{\gamma} \end{array} \right)
                         |\frac{1}{4 m^2 c^2}[J_{\pi}(\vec{\sigma_{j}}.\vec{p_{j}}) - (\vec{\sigma_{j}}.\vec{p_{j}})J_{\pi}(\vec{\sigma_{i}}.\vec{p_{i}}) - (\vec{\sigma_{i}}.\vec{p_{i}})J_{\pi}(\vec{\sigma_{j}}.\vec{p_{j}})\\
                  \nonumber&&    + (\vec{\sigma_{i}}.\vec{p_{i}})(\vec{\sigma_{j}}.\vec{p_{j}})J_{\pi}]|
                         \tilde{\left(
                       \begin{array}{cc}
                         \phi_{\beta} & \phi_{\delta}\\
                        \end{array} \right)}\rangle\\
           \nonumber      &&=  \langle\left(
                       \begin{array}{cc}
                         \phi_{i} & \phi_{j} \end{array} \right)
                         |\frac{1}{4 m^2 c^2}[-J_{\pi}(\vec{\sigma_{j}}.\vec{p})(\vec{\sigma_{i}}.\vec{p}) + (\vec{\sigma_{j}}.\vec{p})J_{\pi}(\vec{\sigma_{i}}.\vec{p}) + (\vec{\sigma_{i}}.\vec{p})J_{\pi}(\vec{\sigma_{j}}.\vec{p})\\
                         \nonumber&&- (\vec{\sigma_{i}}.\vec{p})(\vec{\sigma_{j}}.\vec{p})J_{\pi}]|\tilde{\left(
                       \begin{array}{cc}
                         {\phi_{j}} & {\phi_{i}}\\
                        \end{array} \right)}\rangle
 \end{eqnarray}
Using the relation $(\vec{\sigma_{i}}.\vec{p})(\vec{\sigma_{j}}.\vec{p})= -\hbar^{2}(\vec{\sigma_{i}}.\vec{\sigma_{j}})\frac{d J_{\pi}}{d r}\frac{d}{d r}$ we obtain,
 \begin{eqnarray}
 \nonumber && \langle F_{\alpha}F_{\gamma}|V_{ps}(r)|\tilde{F_{\beta}F_{\delta}}\rangle \\
 \nonumber&&=\langle \phi_{\alpha}\phi_{\gamma}| \frac{1}{4 m^2 c^2}[-J_{\pi}(2(S.\hat{n})^{2}P^{2})+J_{\pi}P^{2}
   -2\hbar^{2}(2S(S+1)-3)\frac{d J_{\pi}}{d r}\frac{d}{d r}\\
   \nonumber&& -2(S.\hat{n})^{2}P^{2}J_{\pi}+p^{2}J_{\pi} ]| \tilde{\phi_{\beta}\phi_{\delta}}\rangle\\
\nonumber &&=\langle \phi_{\alpha}\phi_{\gamma}|\frac{-2 \hbar^{2}c^{2}}{4m^{2}c^{4}}(2S(S+1)-3)(\frac{d J_{\pi}}{d r}\frac{d}{d r})-\frac{\hbar\omega}{2 m c^{2}}(2n+l+\frac{3}{2})\\
\label{moh14}&&(2(S.\hat{n})^{2}-1)J_{\pi}-\frac{\hbar^{2}\omega^{2}}{16 \hbar^{2} c^{2}}(J_{\pi}r^{2})|\tilde{\phi_{\beta}\phi_{\delta}}\rangle
 \end{eqnarray}


\begin{figure}[htp!]
 \includegraphics[width= 8 cm]{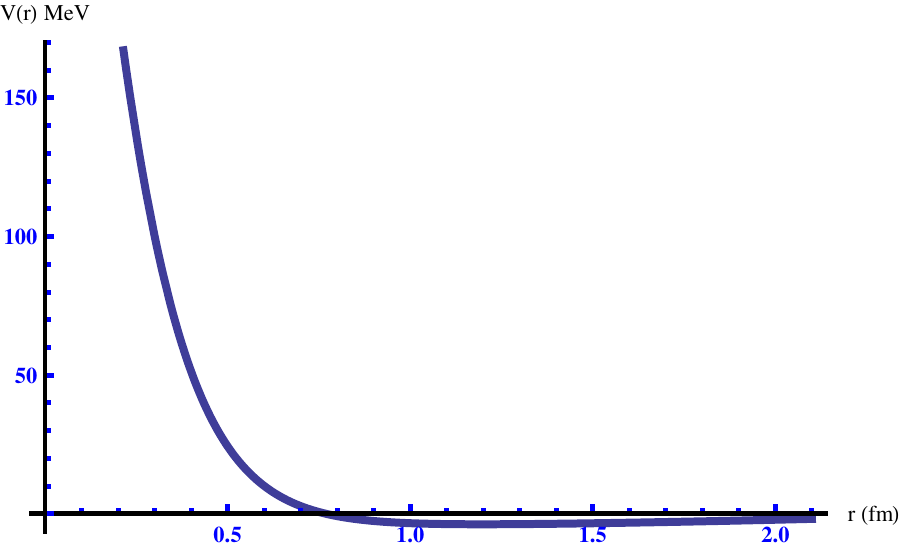},\includegraphics[width= 8 cm]{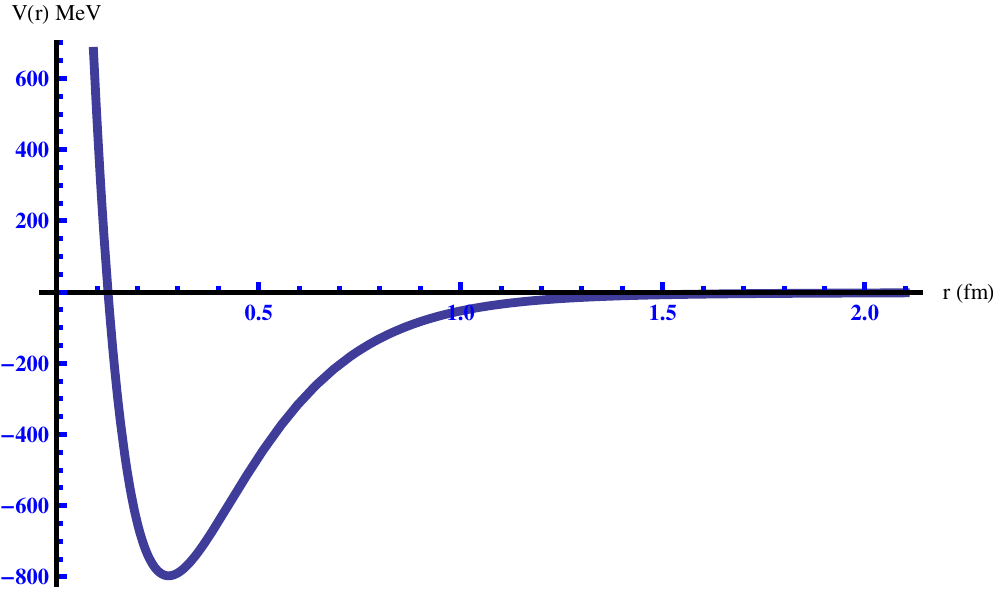}
  \caption{The potential with $GY$ and $SPED$ functions respectively of $^{2}H$ nuclei as OBEP through the exchange of $\sigma$ and $\omega$ mesons , with parameter.I}
  \label{deu1}
\end{figure}
\begin{figure}[p]
  \includegraphics[width= 8 cm]{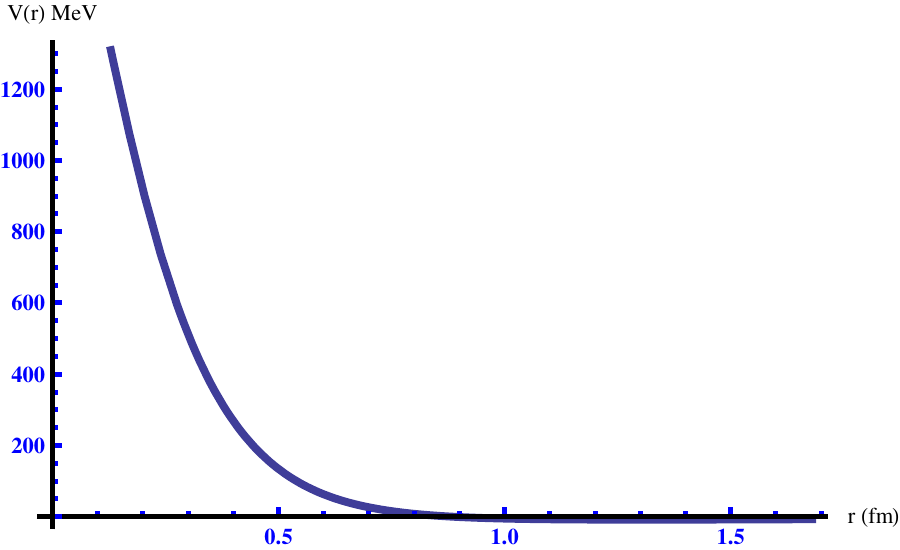},\includegraphics[width= 8 cm]{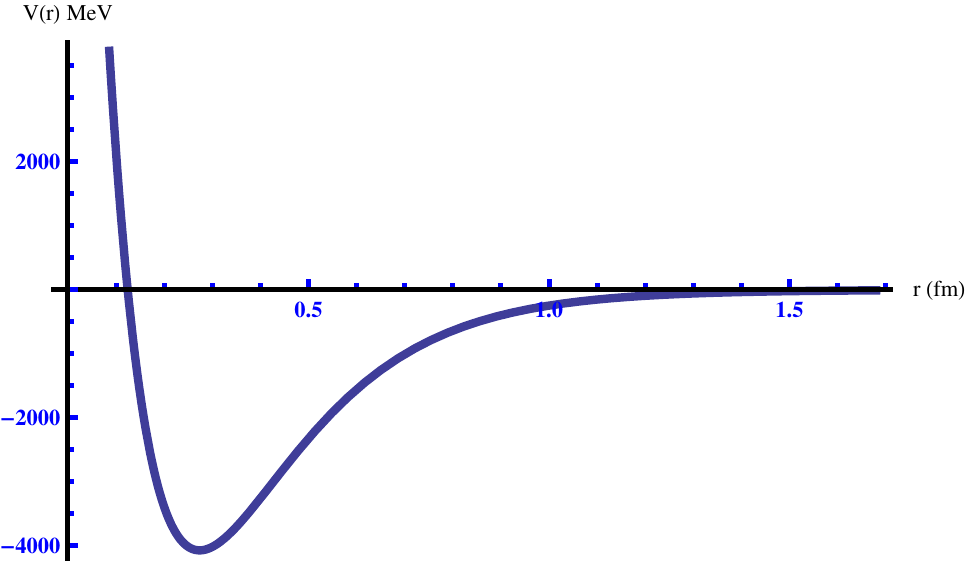}\\
  \caption{The potential with $GY$ and $SPED$ functions respectively of $^{4}He$ nuclei as OBEP through the exchange of $\sigma$ and $\omega$ mesons , with parameter.I}
  \label{deu2}
\end{figure}
\begin{figure}[p]
  \includegraphics[width= 8 cm]{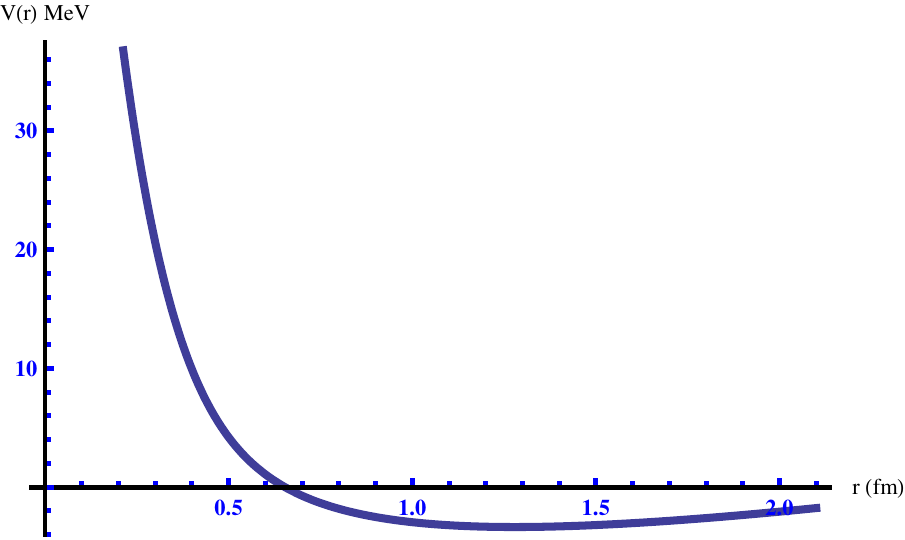},\includegraphics[width= 8 cm]{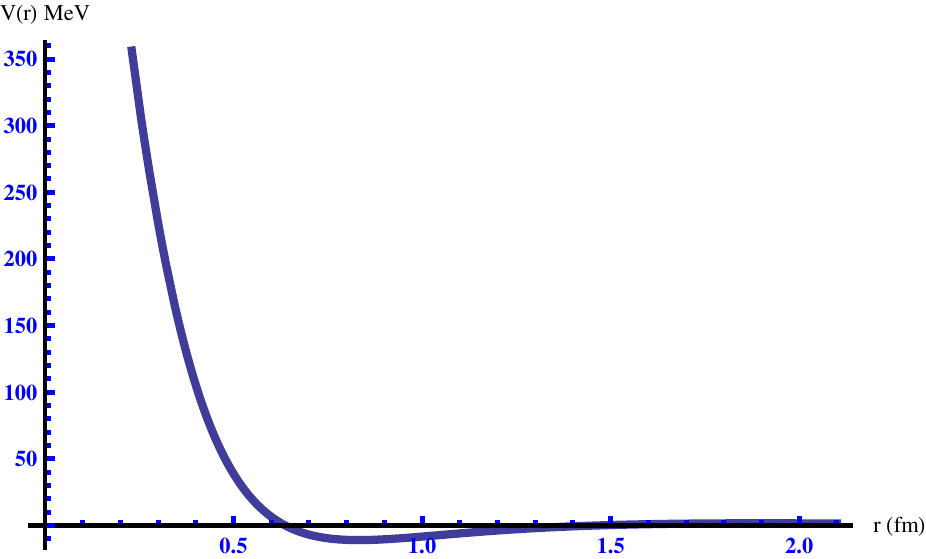}\\
  \caption{The potential with $GY$ and $SPED$ functions respectively of $^{2}H$ nuclei as OBEP through the exchange of $\sigma$ and $\omega$ mesons , with parameter.II}
  \label{deu3}
\end{figure}
\begin{figure}[p]
  \includegraphics[width= 8 cm]{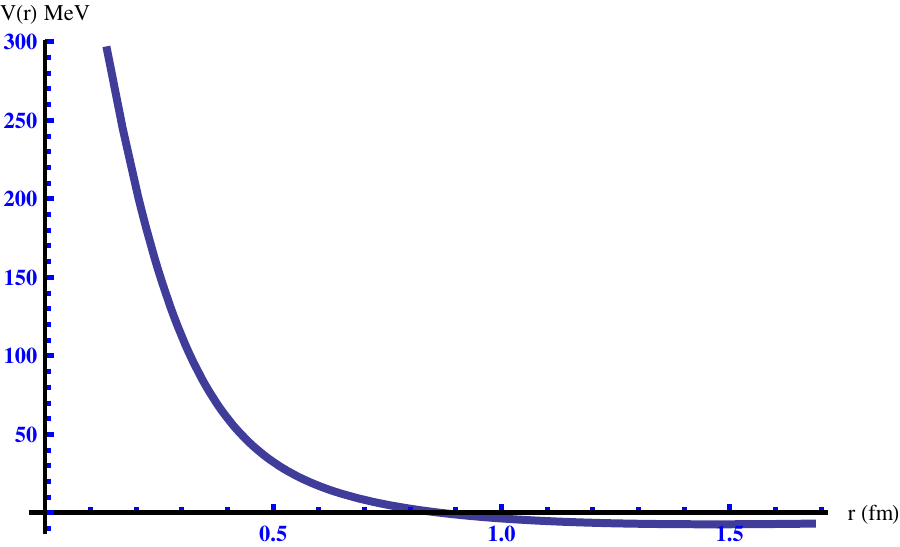},\includegraphics[width= 8 cm]{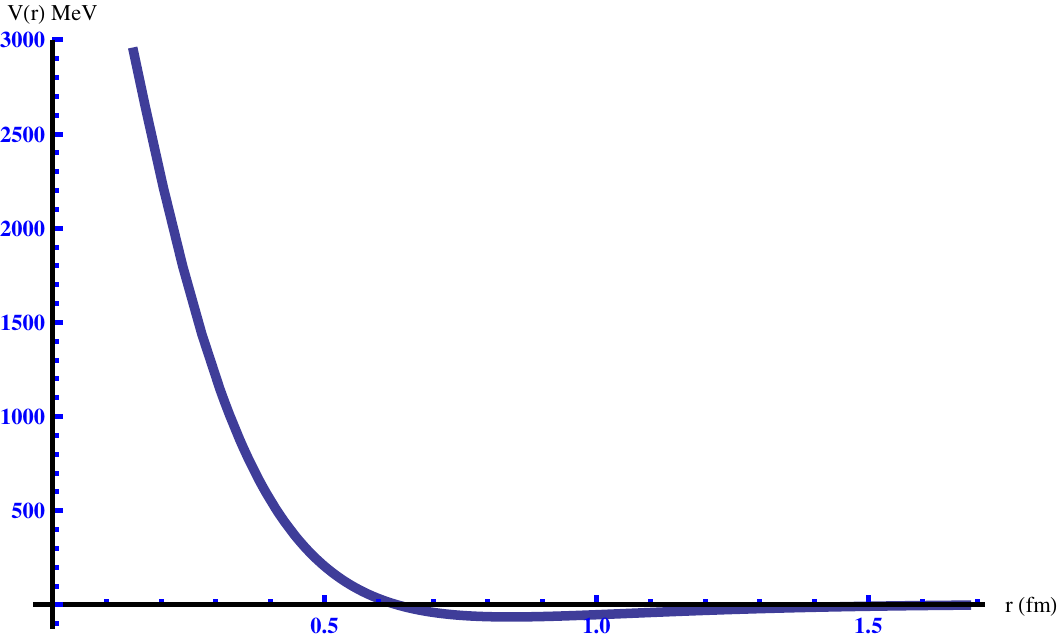}\\
  \caption{The potential with $GY$ and $SPED$ functions respectively of $^{4}He$ nuclei as OBEP through the exchange of $\sigma$ and $\omega$ mesons , with parameter.II}
  \label{deu4}
\end{figure}
\begin{figure}[p]
  \includegraphics[width= 8 cm]{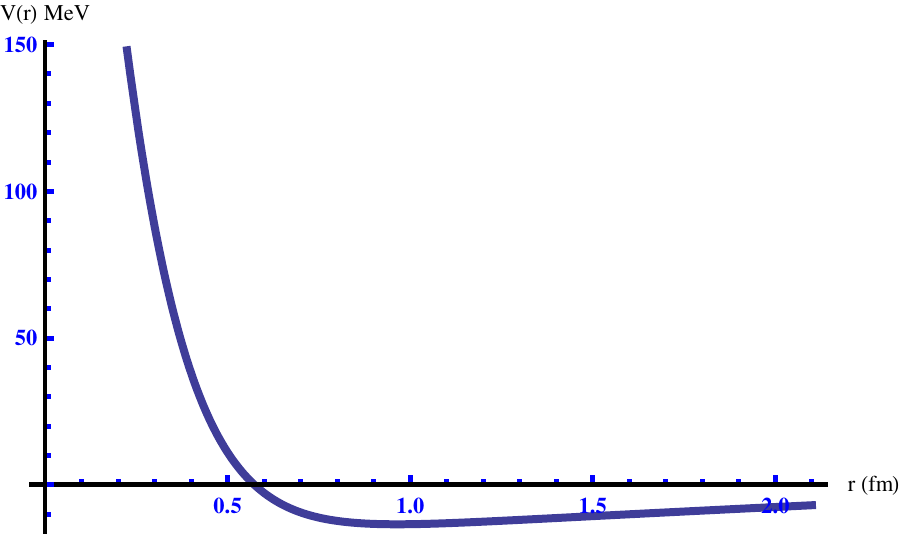},\includegraphics[width= 8 cm]{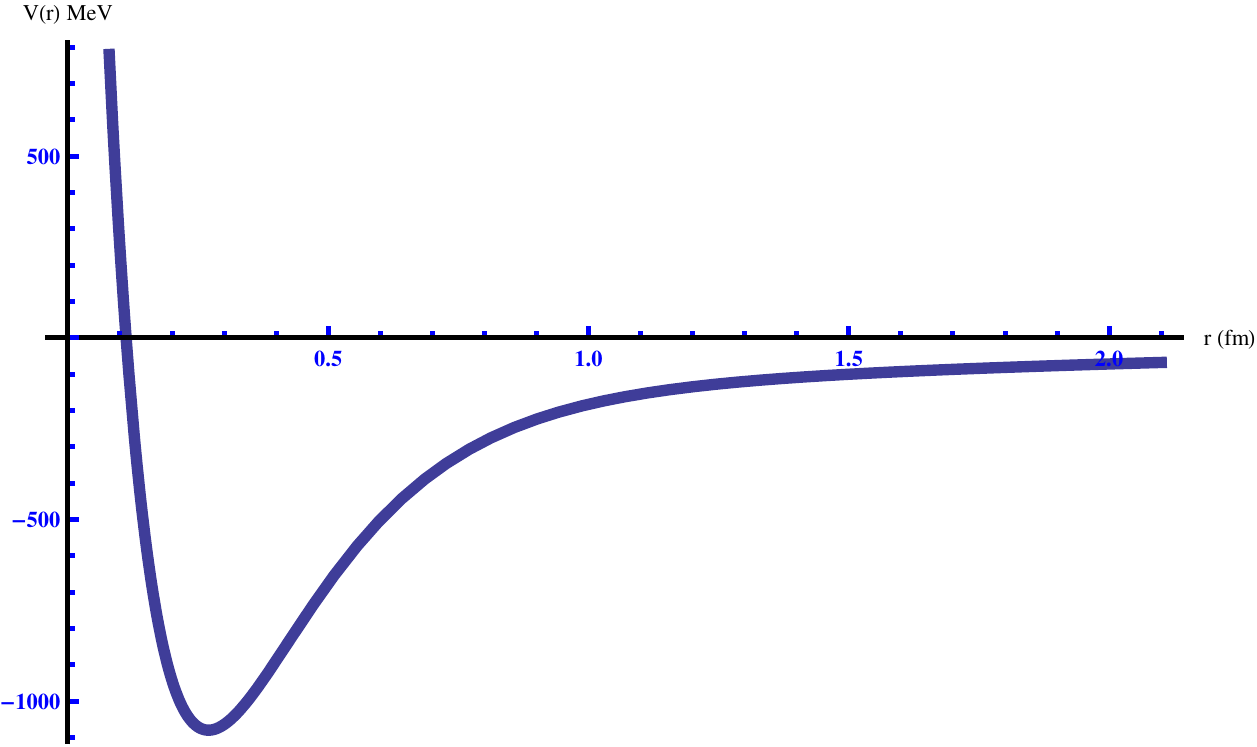}\\
  \caption{The potential with $GY$ and $SPED$ functions respectively of $^{2}H$ nuclei as OBEP through the exchange of $\pi$, $\sigma$ and $\omega$ mesons , with parameter.I}
  \label{deu5}
\end{figure}
\begin{figure}[p]
  \includegraphics[width= 8 cm]{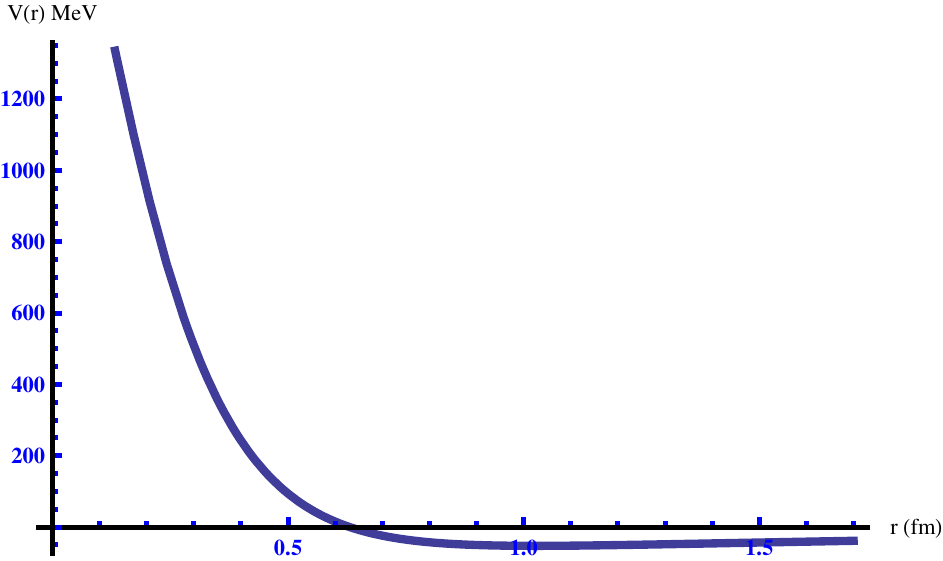},\includegraphics[width= 8 cm]{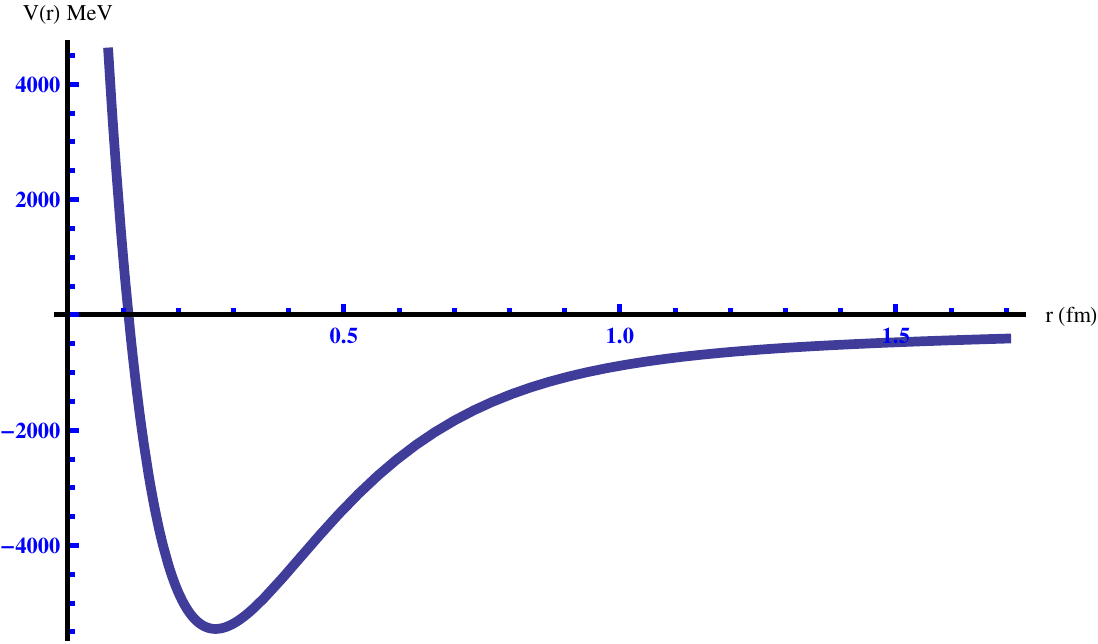}\\
  \caption{The potential with $GY$ and $SPED$ functions respectively of $^{4}He$ nuclei as OBEP through the exchange of $\pi$, $\sigma$ and $\omega$ mesons , with parameter.I}
  \label{deu6}
\end{figure}
\begin{figure}[p]
  \includegraphics[width= 8 cm]{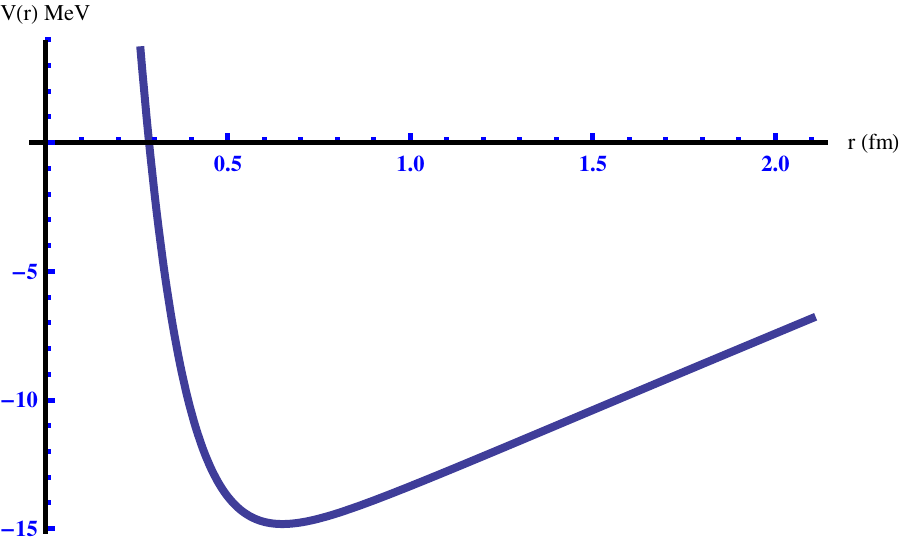},\includegraphics[width= 8 cm]{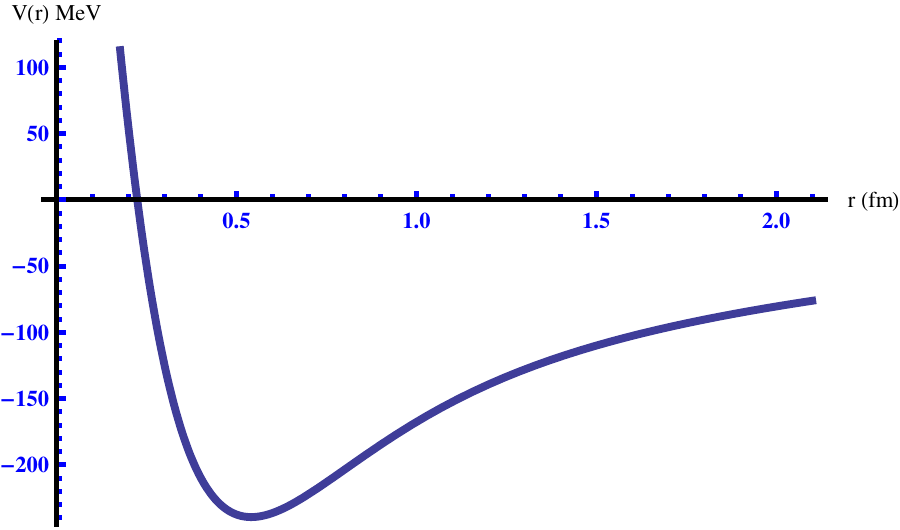}\\
  \caption{The potential with $GY$ and $SPED$ functions respectively of $^{2}H$ nuclei as OBEP through the exchange of $\pi$, $\sigma$ and $\omega$ mesons , with parameter.II}
  \label{deu7}
\end{figure}
\begin{figure}[p]
  \includegraphics[width= 8 cm]{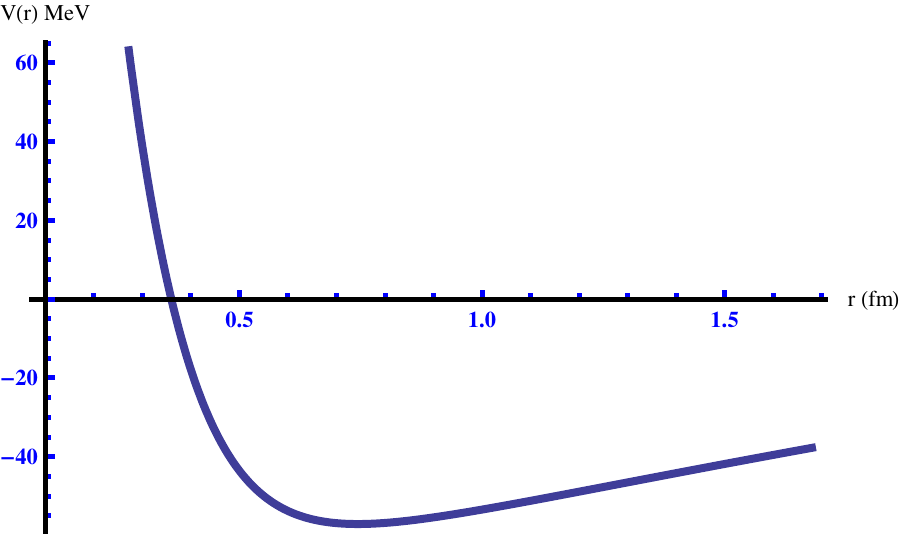},\includegraphics[width= 8 cm]{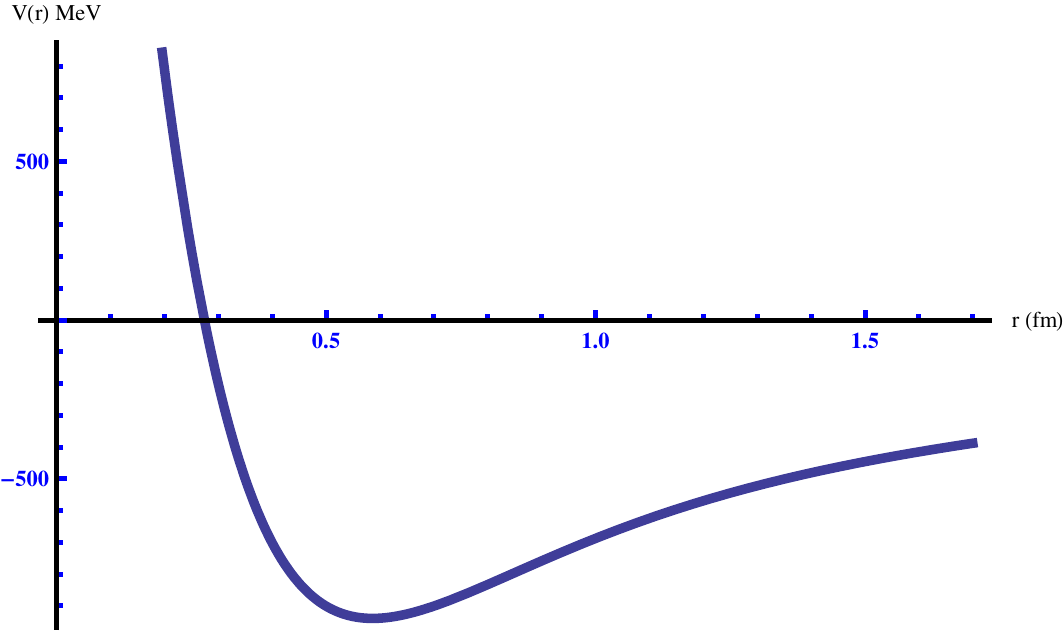}\\
  \caption{The potential with $GY$ and $SPED$ functions respectively of $^{4}He$ nucleus as OBEP through the exchange of $\pi$, $\sigma$ and $\omega$ mesons , with parameter.II}
  \label{deu8}
\end{figure}


\begin{thebibliography}{99}
\bibitem{griffiths} D. Griffiths, Introduction to Elementary Particles\copyright John Wiley and Sons, Inc. (1987).
\bibitem{vallieres}M. Vallieres and H. Wu, The Shell Model\copyright Springer-Verlag Berlin (1991).
\bibitem{kamal85}K. K. Seth, Nucl. Phys. \textbf{A434}, 287(1985).
\bibitem{scot2002}S. Bogner, T. T. S.  Kuo, L. Coraggio, A. Covello, and N. Itaco, Phys. Rev. \textbf{C65}, 051301(R)(2002).
\bibitem{sahu14}B. B. Sahu, S. K. Singh, M. Bhuyan et. al, Phys. rev. \textbf{C89}, 031301(2014).
\bibitem{yukawa35}H. Yukawa, Proc. Phys. Math. Soc. Jpn. \textbf{17}, 48(1935).
\bibitem{james17}J. P. Edwards, U. Gerber, C. Schubert, M. Anabel Trejo, and A. Webery, Prog. Theor. Exp. Phys., \textbf{2017}, 083A01(2017).
\bibitem{taketani52}M. Taketani, S. Machida and S. Ohanuma, Prog. Theor. Phys. \textbf{7}, 45(1952).
\bibitem{brueckner53}K. A. Brueckner and K. M. Watson, Phys. Rev. \textbf{92}, 1023(1953).
\bibitem{signell57}P. S. Signell and R. E. Marshak, Phys. Rev. \textbf{106}, 832(1957).
\bibitem{brown91}G. E. Brown and M. Rho, Phys. Rev. Lett. \textbf{66}, 2720(1991).
\bibitem{naghdi14}M. Naghdi, Phs. Part. Nucl. \textbf{5}, 924(2014).

\bibitem{partovi70}M. H. Partovi and E. L. Lomon, Phys. Rev. \textbf{D2}, 1999(1970).
\bibitem{jackson75}A. D. Jackson, D. O. Riska and B. Verwest, Nucl. Phys. \textbf{A249}, 397(1975).
\bibitem{lacombe73}W. N. Cottingham, M. Lacombe, B. Loiseau, J. M. Richard and R. Vinhman, Phys. Rev. \textbf{D8}, 3633(1973).
\bibitem{nagels78}M. M. Nagels, T. A. Rijken and J. J. Deswart, Phys. Rev. \textbf{D17}, 768(1978).
\bibitem{bouyssy87}A. Bouyssy, J. F. Mothiot, N. Van Giai, and S. Marcos, Phys. Rev. \textbf{C36}, 380(1987).
\bibitem{jaminon81}M. Jaminon, C. Mahaux, and P. Rochus, Nucl. phys. \textbf{A365}, 371(1981).
\bibitem{reid68}R. V. Reid, Ann. Phys. \textbf{50}, 411(1968).
\bibitem{day81}B. D. Day, Phys. Rev. \textbf{C24}, 1203 (1981).
\bibitem{stoks94}V. G. J. Stoks, R. A. M. Klomp, C. P. F. Terheggen and J. J. Deswart, Phys. Rev. \textbf{C49}, 2950(1994).
\bibitem{lagaris81}I. E. Lagaris and V. R. Pandharipande, Nucl. Phys. \textbf{A359}, 349(1981).
\bibitem{wiringa95}R. B. Wiringa, V. G. J. Stoks, R. Schiarilla, Phys. Rev. \textbf{C51}, 38(1995).
\bibitem{lassila62}K. E. Lassila, M. H. Hull, H. M. Ruppel, F. A. Mcdonald and G. Breit, Phys. Rev. \textbf{128}, 830(1962).
\bibitem{hamada62}T. Hamada and I. D. Johnston, Nucl. Phys. \textbf{34}, 382(1962).
\bibitem{watari67}W. Watari, Rev. Mod. Phys. \textbf{39},594(1967).
\bibitem{Machleidt87}R. Machleidt, K. Holinde and Ch. Elster, Phys. Rep. \textbf{149}, 1(1987).

\bibitem{zahra16}Z. Ghalenovi, 19th Iranian Physical Chemistry Conference, 448(2016).
\bibitem{haidenbauer}J. Haidenbauer, Braz. J. phys. \textbf{34}, 846(2004).
\bibitem{miller72}L. D. Miller and A. E. S. Green, Phys. Rev. \textbf{C5}, 241(1972).
\bibitem{meibner2005} U. G. Meibner, Nucl. Phys. \textbf{A751}, 149(2005).
\bibitem{arias2001}J.M. Arias and M. Lozano, An advanced Course in Modern Nuclear Physics, \copyright Springer-Verlag Berlin and Heidelberg (2001).
\bibitem{jaminon80}M. Jaminon, C. Mahaux and P. Rochus, Phys. Rev. \textbf{C22}, 2027(1980).
\bibitem{neff15}T. Neff, H. Feldmeier and W. Horiuchi, Phys. Rev. \textbf{C92}, 024003(2015).
\bibitem{takayuki2009}T. Myo, H. Toki and K. Ikeda, Prog. Theor. Phys. \textbf{121}, 511(2009).
\bibitem{gartenhaus57}S. Gartenhaus and C. Schwaetz, Phy. Rev. \textbf{108}, 482(1957).
\bibitem{wen10}W. H. Long, P. Ring, N. Van Giai and J. Meng, Phys.Rev.\textbf{C81}, 024308(2010).
\bibitem{stakgold}I. Stakgold, Green's Functions and Boundary Value Problems \copyright Wiley, New York(1979).
\bibitem{lebeddev}N. Lebedev, Special Functions and Their Applications \copyright Prentice-Hall, Englewood Cliffs, NJ,(1965).
\bibitem{Kirson}M. W. Kirson, Nucl. Phys. \textbf{A }, 781(2007).
\bibitem{Kenneth}K. S. Krane, Introductory nuclear physics \copyright John Wiley  Sons, Inc.(1988).
\bibitem{green67}A. E. S. Green, T. Sawada, Rev. Mod. phys. \textbf{39}, 594(1967).
\bibitem{anselm99}A. Anselm and N. Dombey, J. Phys. G:Nucl. Part. Phys. \textbf{25}, 513(1999).
\bibitem{franz92}F. Gross, J. W. Van Orden and K. Holinde, Phys. Rev. \textbf{C45}, 2094(1992).
\bibitem{Jacques2000}K. A. Amos, P. J. Dortmans, H. V. von Geramb, S. Karataglidis, J. Raynal, Adv. Nucl. Phys. \textbf{25}, 276(2000).
\bibitem{Varshalovich88}D. A. Varshalovich, A. N. Moskalev, V. K. Khersonski, Quantum Theory of Angular Momentum \copyright World Scientific Co.Pte.Ltd.(1988).
\bibitem{Tyren66}H. Tyren, S. Kullander, O. Sundberg, R. Ramachandran, P. Isacsson, T. Berggren, Nucl. Phys \textbf{79}, 321(1966).
{\color{blue}\bibitem{gad2011}K. Gad, Ann. phy. \textbf{326}, 2474(2011).}
\bibitem{abdo} J. W. Negele and E. Vogt, Advances in nuclear physics 19 \copyright Plenum Press New York-London(1989).
{\color{blue}\bibitem{haung}F. Huang and W. L. Wang, Phys. Rev. D 98, 074018 (2018).}
\bibitem{shehab} O. Shehab, K. Landsman, Y. Nam, D. Zhu, N. M. Linke, M. Keesan, R. C. Pooser, and C. Monroe, phy. rev. \textbf{A 100}, 062319(2019).
\bibitem{exp}M. Garcona and J.W. V. Orden, DAPNIA/SPHN-01-02, JLAB-THY-01-6(2001).
\bibitem{kessler} G. L. Greene, E. G. Kessler, Jr., R. D. Deslattes, and H. Börner, Phys. Rev. Lett. \textbf{56}, 819(1986).
\bibitem{deuter}E. G. Kessler, M. S. Deweya, R. D. Deslattesa, A. Heninsa, H. G. Börnerb, M. Jentschelb, C. Dollb, H. Lehmannb, Phys. Lett. \textbf{A255}, 221(1999).
\bibitem{binding} //www.nndc.bnl.gov/nudat2/recenter.jsp,z=1,n=1.
\bibitem{kowalski} K. Kowalski, D. J. Dean, M. H. Jensen, T. Papenbrock, and P. Piecuch, Phy. Rev. Lett. \textbf{92}, 13(2004).
\bibitem{helium}https://www.nndc.bnl.gov/nudat2/getbandplot.jspnucleus=4HE.

\bibitem{Wiringa} R. B. Wiringa, Phys. Rev. \textbf{C43}, 1585(1991).
\bibitem{Iyad18} I. Alhagaish, A. Abu-Nada, F. Afaneh, and M. Hasan, arXiv:nucl-th/1803.04650v1.
\bibitem{rois} J. W. T. Keeblea, A. Rios, https://arxiv.org/abs/1911.13092(2019).
\bibitem{Raynal} J. Raynal, arXiv:nucl-th/0407060(2004).
\bibitem{Gad}K. Gad, Ann. of Phys., \textbf{327}, 2403(2012).

\end{thebibliography}
\end{document}